\documentclass[conference]{IEEEtran}
\IEEEoverridecommandlockouts
\usepackage{amsmath,amssymb,amsfonts}
\usepackage{algorithmic}
\usepackage{graphicx}
\usepackage{textcomp}
\usepackage{xcolor}
\usepackage{colortbl}
\usepackage{todonotes}
\usepackage{adjustbox}
\usepackage{tabularx}
\usepackage{booktabs}
\usepackage{caption}
\usepackage{subcaption}

\usepackage[capitalize]{cleveref} 

\def\BibTeX{{\rm B\kern-.05em{\sc i\kern-.025em b}\kern-.08em
    T\kern-.1667em\lower.7ex\hbox{E}\kern-.125emX}}
\usepackage[style=numeric,sorting=none]{biblatex}
\begin{document}

\title{Autonomy and Intelligence \\in the Computing Continuum: \\ Challenges, Enablers, and Future Directions for Orchestration
\thanks{This research has been supported by the Academy of Finland, 6G Flagship program under Grant 346208; the ECSEL JU FRACTAL (grant 877056), receiving support from the EU Horizon 2020 programme and Spain, Italy, Austria, Germany, France, Finland, Switzerland; and the Infotech Oulu research institute.}
}

\author{\IEEEauthorblockN{
    Henna Kokkonen\IEEEauthorrefmark{1}\IEEEauthorrefmark{8},
    Lauri Lov{\'e}n\IEEEauthorrefmark{1}\IEEEauthorrefmark{9}\IEEEauthorrefmark{8}, 
    Naser Hossein Motlagh\IEEEauthorrefmark{6},
    Abhishek Kumar\IEEEauthorrefmark{1},
    Juha Partala\IEEEauthorrefmark{2},
    \\
Tri Nguyen\IEEEauthorrefmark{1},
    Víctor Casamayor Pujol\IEEEauthorrefmark{9},
    Panos Kostakos\IEEEauthorrefmark{1},\\
    Teemu Leppänen\IEEEauthorrefmark{1},
    Alfonso Gonz{\'a}lez-Gil\IEEEauthorrefmark{7},
    Ester Sola\IEEEauthorrefmark{7},
    Iñigo Angulo\IEEEauthorrefmark{7},
    Madhusanka Liyanage\IEEEauthorrefmark{4}\IEEEauthorrefmark{5},\\
    Mehdi Bennis\IEEEauthorrefmark{4},
    Sasu Tarkoma\IEEEauthorrefmark{6}\IEEEauthorrefmark{1},
    Schahram Dustdar\IEEEauthorrefmark{9},
    Susanna Pirttikangas\IEEEauthorrefmark{1},
    Jukka Riekki\IEEEauthorrefmark{1}
}
\IEEEauthorblockA{
    \IEEEauthorrefmark{1}Center for Ubiquitous Computing, University of Oulu, Finland\\
    \IEEEauthorrefmark{2}Center for Machine Vision and Signal Analysis, University of Oulu, Finland\\
    \IEEEauthorrefmark{4}Center for Wireless Communications, University of Oulu, Finland\\   
    \IEEEauthorrefmark{5}School of Computer Science, University College Dublin, Ireland\\
    \IEEEauthorrefmark{6}Department of Computer Science, University of Helsinki, Finland\\
    \IEEEauthorrefmark{9}Distributed Systems Group, TU Wien, Austria\\
    \IEEEauthorrefmark{7}R\&D Department, Zylk.net S.L., Spain\\
    Email: \IEEEauthorrefmark{1}\IEEEauthorrefmark{2}\IEEEauthorrefmark{4}\{firstname.lastname\}@oulu.fi,
    \IEEEauthorrefmark{5}madhusanka@ucd.ie}\\
    These authors contributed equally: \IEEEauthorrefmark{8}
}

\maketitle

\begin{abstract}
Future AI applications require performance, reliability, and privacy that the existing, cloud-dependant system architectures cannot provide. In this article, we study orchestration in the device-edge-cloud continuum, and focus on edge AI for resource orchestration. We claim that to support the constantly growing requirements of intelligent applications in the device-edge-cloud computing continuum, resource orchestration needs to embrace edge AI and emphasize local autonomy and intelligence.
To justify the claim, we provide a general definition for continuum orchestration, and look at how current and
emerging orchestration paradigms are suitable for the computing continuum. We describe certain major emerging research themes that may affect future orchestration, and provide an early vision of an orchestration paradigm that embraces those research themes. Finally, we survey current key edge AI methods and look at how they may contribute into fulfilling the vision of future continuum orchestration.
\end{abstract}


\section{Introduction}\label{sec:Introduction}
Bringing intelligence \emph{to} the edge requires intelligence \emph{on} the edge. We argue that the future computing continuum, spanning from devices and edge/fog nodes to the cloud, cannot be seen as a static environment. Moreover, orchestration in the continuum cannot rely on the traditional best-effort, reactive, threshold based methods of the current centralized orchestration paradigms developed for the cloud. Instead, the continuum must embrace autonomy and distributed intelligence.

In more detail, future AI applications require performance, reliability and privacy that the existing, cloud-dependant system architectures cannot provide. In these systems, sensor data is transported to the cloud for AI analytics, overburdening the networks, consuming a lot of energy, increasing the latency of task processing, and raising privacy concerns. These issues have caused a paradigm shift: instead of having all AI processing in the cloud, the intelligence is brought to the edge, closer to the data generating devices, users, and controlled devices. 

\begin{figure}[t]
    \centering
    \includegraphics[width=\linewidth]{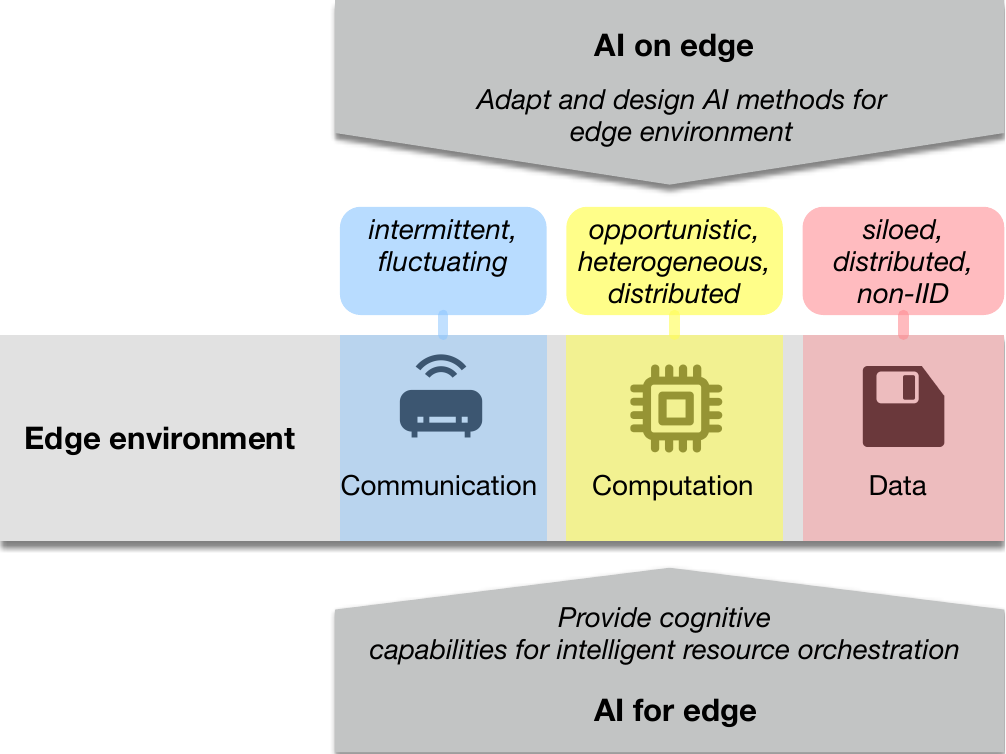}
    \caption{AI on edge vs. AI for edge.}\label{fig:edge_environment}
\end{figure}

However, while edge-based processing increases data privacy and shortens latencies, deploying AI applications on the edge introduces significant challenges. Edge is a distributed platform of heterogeneous nodes, characterized by fluctuating and intermittent communication, opportunistic, heterogeneous and distributed computational resources, and siloed, distributed, and non-IID data (\cref{fig:edge_environment}). Transferring centralized AI with high resource requirements to the distributed, resource-constrained environment, while providing guarantees for performance, reliability, and security, requires new architectures and algorithms for training, inference and decision making over wireless and fixed links.

On the other hand, edge/fog computing is based on the deployment of a wide variety of applications on heterogeneous hardware components (e.g., CPU, GPU, or FPGA), each with its own software tools and dependencies. To mitigate the heterogeneity of the platforms, virtualization techniques are widely used to extend the physical limitations of the infrastructure, as they provide isolated software environments and full abstraction of the software or hardware running behind the environment. This includes IT approaches, such as virtual machines (VM) and containers \cite{Tao2019}, which share a pool of resources that can be scaled dynamically, and network approaches, such as Network Function Virtualization (NFV), Software Defined Networking (SDN), and Network Slicing \cite{Taleb2017,Afolabi2018}, which permit decoupling the data and control planes from the constraints of the physical network. This allows multi-tenant environments to adapt to different edge/fog computing scenarios.

These enabling technologies lead towards a wide range of new and innovative services \cite{Taleb2017}. Many of them, such as computation offloading, multi-site collaboration and context awareness, are presented as disruptive approaches that will change current computing paradigms, fulfilling, for example, the needs of distributed and cooperative AI applications \cite{dafoe2020open,Dafoe2021}. As the computational workloads of these services become more demanding, automating their deployment and management over the available computational resources is a requirement for efficient use of resources and optimal performance.

Furthermore, to reach the device-edge-cloud computing continuum envisioned as the future direction in 5G and beyond architectures \cite{Taleb2017, Samdanis2020B5G}, telecommunication and IT services must convergence and be jointly optimized. Physical infrastructure, comprising IT elements (e.g., machine clusters in the cloud or edge/fog servers) and network components (access points and transport networks), has to cater for the needs of an opportunistic computing continuum, provisioning services flexibly and scalably \cite{Zhao2019}.

Increasing the intelligence of the edge platform itself, promoting next generation services, and bringing about a coherent computing continuum are therefore necessities for supporting the AI workflows on the edge. To this end, 5G technology has taken initiative in improving the reliability and scalability of wireless networks by introducing Ultra-reliable low-latency communication (URLLC), Enhanced Mobile Broadband (eMBB), and Massive Machine Type Communication (mMTC). Furthermore, reaching the 6G vision of ubiquitous wireless intelligence and use cases in, e.g., extended reality (XR) and telepresence \cite{aazhang2019} or massive environmental sensing \cite{su2021intelligent,loven2021edison} requires autonomous, context-aware, self-learning and self-optimising computational, communicational and data storage services that will fulfill the growing requirements of security, efficiency, and reliability. 

Subsequently, the device-edge-cloud computing continuum, as a crucial element of the next generation mobile and fixed networks, requires AI solutions for its orchestration, that is, for managing, allocating and distributing its resources in a dynamic, efficient, and context-aware manner.

Designing AI methods for the edge environment and deploying AI applications there, and developing AI solutions that intelligently configure, control and manage the edge environment to optimize its resource usage, have respectively been coined as \emph{AI on edge} and \emph{AI for edge} \cite{Deng2019,loven2019}. Together, these aspects form \emph{edge AI} or \emph{edge intelligence} (\cref{fig:edge_environment}). 

This confluence of edge computing and AI with the next generation of communication and computation in the device-edge-cloud continuum has been envisioned as a fundamental key factor in the creation of the \emph{Intelligent Internet of Intelligent Things} \cite{peltonen2020}. The culmination of this confluence is distributed, edge-native AI, which will enable novel, smart applications in domains such as urban computing, smart environments, personalized services and context-aware mobile technologies \cite{loven2019}, supporting sustainability and resilience through efficient resource usage and dynamic and context-aware operation.

\subsection{Contributions}

Taking a very different point of view compared to related work (\cref{sec:Related}), we focus on holistically bringing together distributed AI and orchestration in the device-edge-cloud computing continuum. We adopt the Multi-Agent System (MAS) paradigm to regard the edge platform as a dynamic network of AI nodes. Each node is a highly adaptive, self-interested, autonomous and context-aware agent that is able to monitor the state of the surrounding dynamic and complex edge environment, forecast internal and external events, and react proactively to changes based on the forecasts. We build the survey around a vision for AI for edge, that is, by developing AI solutions for the orchestration of the network, the edge environment will eventually evolve into a coherent computing continuum that is able to function in an autonomous, decentralized and decoupled manner, while optimizing and balancing multiple objectives with regard to, e.g., efficiency, reliability and security. The network will be able to orchestrate its limited computational, network, energy and memory resources in a globally optimized manner while being aware of and ready to adapt to the dynamic environment.

The overall contribution of this survey is to justify the claim that to support the constantly growing requirements of intelligent applications in the device-edge-cloud computing continuum, resource orchestration needs to embrace edge AI and emphasize local autonomy and intelligence. In more detail, the contribution can be broken down as follows:
\begin{itemize}
    \item We provide an encompassing view on computing continuum resources and orchestration, proposing a broader definition for orchestration, which relies on the finding that considering \emph{Everything as a Resource} (EaaR) simplifies the overall view.
    \item We bring together several different orchestration taxonomies from the literature, proposing a new, more holistic taxonomy.
    \item We introduce current and emerging orchestration paradigms, pinpointing their main deficiencies with regard to achieving an autonomous and intelligently optimized device-edge-cloud computing continuum orchestration.
    \item We provide an early vision of autonomous, weakly coupled and secure orchestration in the computing continuum.
    \item We identify the main opportunities and promising research directions brought forth by the vision, and identify the main challenges that currently obstruct its realization. 
    \item We provide an extensive overview of the current state of the art in AI methods that we regard as the key factors in realizing an autonomous computing continuum.
\end{itemize}

The rest of the article is structured as follows. \cref{sec:Related} gives an overview of related work in edge AI and orchestration. \cref{sec:Definitions} briefly defines the main concepts required for studying orchestration, that is, agent modelling, distributed AI, autonomy, computing continuum and orchestration, and sets the scope using these definitions. \cref{sec:Edgeorchestration} presents current and emerging orchestration paradigms, while \cref{sec:Vision} charts out the opportunities and development directions in computing continuum orchestration, and presents an early vision of autonomous orchestration in the computing continuum. 
\Cref{sec:Enablers} identifies requirements for AI methods in the computing continuum, and surveys the current state of the art, mainly focusing on methods in distributed learning and decision making, while also covering some other promising approaches. \Cref{sec:Discussion} illustrates how edge AI can potentially be used in the computing continuum orchestration. Finally, Section \ref{sec:Conclusion} concludes the article.

\section{Related Work}\label{sec:Related}
\begin{table*}
\centering
\caption{Edge Intelligence related surveys}
\begin{tabular}{p{0.14\textwidth}p{0.08\textwidth}p{0.34\textwidth}p{0.34\textwidth}} 
 \toprule
 \textbf{Reference} & \textbf{Topic} & \textbf{Main contribution} & \textbf{Limitations} \\
 \midrule
 Deng et al. 2019 \cite{Deng2019} & Edge AI & Surveys compactly both sides of Edge AI, that is, AI for Edge and AI on Edge. Provides a taxonomy for AI for Edge, which consists of three categories: Topology, Content and Service. & State of the art in Edge for AI focuses on Wireless Networking, Service Placement and Caching, and Offloading.
 \\
 \midrule
 Xu et al. 2020 \cite{Xu2020} & Edge AI & Focuses on AI on Edge aspect, categorizes Edge intelligence into Edge Training, Edge Inference, Edge Caching and Edge Offloading. & Does not survey AI methods that have been applied to solving edge problems. \\
 \midrule
 Lim et al. 2020 \cite{Lim2019} & FL & Focuses on the FL implementation challenges and solutions on edge, as well as surveys how FL has been used as a tool in optimizing mobile edge network problems. & Focuses solely on the application of FL in MEC problems. \\
  \midrule
 Park et al. 2019 \cite{Park2019} & Edge ML & Focuses on technical and theoretical enablers of Edge ML, particularly from the viewpoint of efficient communication. & Main emphasis is on AI on Edge aspect, i.e., enablers for training an accurate model on edge with low latency, high scalability and efficiency. \\
 \midrule
 Park et al. 2021 \cite{Park2021} & Edge ML & Surveys communication and ML principles and enablers for achieving communication-efficient distributed learning on edge. & Focuses on AI on Edge aspect. \\
 \midrule
 Wang et al. 2020 \cite{Wang2020} & Edge DL & Surveys comprehensively the relationship of DL and Edge Computing from the viewpoints of DL for Edge and DL on Edge. & Focuses solely on DL techniques. \\
 \midrule
 Zhou et al. 2019 \cite{Zhou2019} & Edge AI & Surveys architectures, key performance indicators and enabling technologies for DL model training and inference on edge. & Focuses on AI on Edge aspect. \\
 \midrule
 Hao et al. 2021 \cite{Hao2021} & Edge AI & Surveys enabling single-layer and cross-layer design methodologies for Edge AI. & AI on Edge viewpoint, that is, what kind of hardware and software design methodologies exist for developing and deploying AI applications on Edge. \\
 \midrule
 Pham et al. 2020 \cite{Pham2020} & MEC & As a part of the survey, how ML has been applied to solving MEC problems is covered. & Provides a concise overview of the state of the art in each MEC problem area, focusing on DL based works. \\
 \midrule
 Xu et al. 2022 \cite{Xu2022iov} & AI for IoV & Surveys AI methods that have been utilized in optimizing edge server placement, as well as computation and service offloading in Internet of Vehicles. & Focus on IoV. \\
 \midrule
 Shi et al. 2020 \cite{Shi2020} & Edge AI & Surveys communication challenges and proposed communication-efficient algorithms and systems for Edge AI. & Focuses on AI on Edge aspect (edge training and edge inference).\\
 \midrule
 \bottomrule
\end{tabular}
\label{table:relatedsurveysAI}
\end{table*}

\begin{table*}
\centering
\caption{Orchestration related surveys}
\begin{tabular}{p{0.14\textwidth}p{0.08\textwidth}p{0.34\textwidth}p{0.34\textwidth}} 
 \toprule
 \textbf{Reference} & \textbf{Topic} & \textbf{Main contribution} & \textbf{Limitations} \\
 \midrule
 Taleb et al. 2017 \cite{Taleb2017} & MEC orchestration & Provides a survey for MEC service orchestration in edge-cloud architectures. & The focus is on service orchestration in edge-cloud continuum. \\
 \midrule
 Guerzoni et al. 2017 \cite{guerzoni2017analysis} & Multi-domain orchestration & Provides a survey on architectures for orchestration in software defined infrastructures and proposes a reference architecture for end-to-end multi-domain orchestration. & Focuses on architectures for enabling multi-domain resource and service orchestration in software defined networks, lacks an edge computing aspect, and no consideration of AI based orchestration solutions. \\
 \midrule
 Tocz{\'e} and Nadjm-Tehrani 2018 \cite{tocze2018taxonomy} & Resource management & Proposes a taxonomy for resource management in the device-edge-cloud continuum. & The focus is on the management of physical and virtual resources. \\
 \midrule
 de Sousa et al. 2019 \cite{de2019network} & Network orchestration & Provides a comprehensive survey of network service orchestration and proposes a taxonomy for network orchestration approaches.
 & Focuses on the life cycle management of network services and does not include an edge computing aspect. \\
 \midrule
 Hong and Varghese 2019 \cite{hong2019resource} & Resource management & Proposes a taxonomy for resource management in fog computing, focusing on architectures, infrastructure and algorithms. & The focus is on the management of physical and virtual resources. \\
 \midrule
 Zhong et al. 2021 \cite{Zhong2021} & Container orchestration & Provides a comprehensive overview of and a taxonomy for ML based container orchestration. & Focuses on container life cycle management. \\
 \midrule
 Versluis and Iosup 2021 \cite{Versluis2021} & Workflow orchestration & Provides a systematic literature review on workflow orchestration and taxonomizes four key areas inside it. & The focus is on the management of workflows. \\
 \midrule
 Mampage et al. 2022 \cite{mampage2021holistic} & Resource management & Provides a taxonomy for resource management in serverless computing environments. & The focus is on the management of physical and virtual resources. \\
 \midrule
 Costa et al. 2022 \cite{costa2022orchestration} & Fog orchestration & Proposes a generic architecture for fog orchestration based on an extensive literature review. & The architecture consolidates approaches from the literature, that is, the paper does not state concrete enablers for realising the whole architecture. \\
 \midrule
 \bottomrule
\end{tabular}
\label{table:relatedsurveysOrch}
\end{table*}
\subsection{Edge AI}
Recently, there has been a surge in edge AI related surveys. As seen in \cref{table:relatedsurveysAI}, which summarizes the main contributions and limitations of such surveys, the majority of them strongly emphasize the AI on edge aspect. In other words, the focal point of interest has been in how to enable AI model learning on top of the edge infrastructures. For example, Park et al. \cite{Park2019} provide a very comprehensive survey of communication-efficient model building on edge, focusing on introducing technical and theoretical enablers for accurate and low latency model training and inference.

We focus on AI for edge. This aspect of applying AI techniques to the optimization of edge in order to make it function in a more intelligent and autonomous manner has received considerably less attention. Most notably, Deng et al. \cite{Deng2019} categorize AI for edge into Topology (Orchestration of Edge Sites and Wireless Networking), Content (Data Provisioning, Service Provisioning, Service Placement, Service Composition and Service Caching) and Service (Computation Offloading, User Profile Migration and Mobility Management). However, they do not provide a comprehensive overview of different AI methods utilized inside this taxonomy, but focus only on what AI methods have been applied to Wireless Networking, Service Placement and Caching, and Offloading.

On the other hand, Lim et al. \cite{Lim2019} focus on how Federated Learning (FL) has been applied to edge network problems, namely in cyberattack detection, edge caching and offloading, base station association and vehicular networks. As a part of their Multi-access Edge Computing (MEC) survey, Pham et al. \cite{Pham2020} investigate how Machine Learning (ML) has been applied to several different edge problems: edge caching, computation offloading, resource allocation, security and privacy, big data analytics, and mobile crowdsensing. However, in both articles, the overview is limited, in the former due to the focus only on FL, and in the latter because the authors focus on providing a very concise overview of ML for edge in the context of a broader survey on MEC.

Wang et al. \cite{Wang2020} provide a comprehensive survey of the union of Deep Learning (DL) and Edge Computing. They focus mainly on DL on Edge aspect, but also give an overview of how DL has been used for optimizing edge. The covered edge management issues are edge caching, task offloading, edge communication, security and joint edge optimization. Finally, Xu et al. \cite{Xu2022iov} survey AI solutions for edge server placement and offloading in Internet of Vehicles (IoV). Their survey is very limited due to the focus on IoV.

\subsection{Orchestration}\label{sec:RelWork_Orch}
The device-edge-cloud continuum contains many elements, from fundamental ones such as energy or time through hardware and virtual elements to service, application and workflow related elements. In literature, there is an abundance of orchestration surveys that usually focus on a specific subset of these elements, as well as on a specific part of the device-edge-cloud continuum. In other words, a majority of the surveys focus on a specific domain, such as the management of containers, networks or tasks. There does not seem to exist surveys that would try to holistically piece together different aspects on orchestration in the whole device-edge-cloud continuum. Hence, in this section, we will highlight significant surveys from different aspects. These surveys serve as the main basis for our more encompassing view to orchestration in the device-edge-cloud continuum. The main contributions and limitations of the presented surveys are summarized in \cref{table:relatedsurveysOrch}.

Tocz{\'e} and Nadjm-Tehrani \cite{tocze2018taxonomy}, Hong and Varghese \cite{hong2019resource}, as well as Mampage et al. \cite{mampage2021holistic} all provide surveys on resource management. Tocz{\'e} and Nadjm-Tehrani propose a taxonomy for resource management at the edge. Their taxonomy has four main categories: resource type, resource management objective, resource location, and resource use. They name six different resource types: computation, communication, storage, data, energy and generic. 
For the objective of the resource management, they name five main goals: estimation, discovery, allocation, sharing, and optimization. The resource location encompasses two perspectives: where the resources are located within the architectures, and whether the resources are stationary or mobile. 
Finally, the resource use encompasses functional and nonfunctional properties, which characterize the purpose for which the resource will be used.

Hong and Varghese propose a taxonomy for resource management in fog/edge computing, focusing on three main aspects: architectures, infrastructure, and algorithms. First, architectures are classified based on data flow, control and tenancy. 
Data flow architectures can be divided into aggregation, sharing between peers, and offloading to edge. Control of resources can be centralized, distributed or hierarchical. Tenancy relates to whether an edge node must host multiple applications and support multiple users. Second, infrastructure is classified into hardware (computing and network devices), software (manages the resources of the devices) and middleware (complementary services to system software). 
Finally, the resource management algorithms are classified into discovery, benchmarking, load-balancing, and placement.

Finally, Mampage et al. propose a taxonomy for resource management in serverless environments. They consider three major aspects to resource management in serverless platforms: application workload modelling, resource scheduling and resource scaling. The taxonomy has four main categories: resource management elements, which consists of the three aforementioned main aspects; deployment environment, which relates to understanding the factors influencing the design of a serverless platform; workload management, which relates to understanding the application requirements and structure, the nature of the workload, and data locality; and Quality of Service (QoS) goal.

The viewpoint of all three resource management surveys is limited to the orchestration of the lower level elements, i.e., fundamental, physical and virtual elements.

Taleb et al. \cite{Taleb2017} provide a survey on MEC orchestration, focusing in particular on MEC service orchestration while also providing insights into the joint orchestration of MEC services and virtualized network functions. They have three main aspects on MEC orchestration: 
(1) resource allocation, service placement, platform selection for a service request, and analyzing the reliability of MEC service deployments; (2) supporting service mobility; (3) the joint optimization of virtualized network functions and MEC services on platforms. Their view is limited on the orchestration of services and virtualized network functions on edge-cloud platforms.

Guerzoni et al. \cite{guerzoni2017analysis} survey different architectures for end-to-end orchestration of resources and services in the future 5G environments. They also present a reference architecture that includes all the capabilities that end-to-end orchestration is expected to fulfill. They name multi-domain orchestration as the key enabler behind end-to-end orchestration that spans from the infrastructure layer to the application layer. 
They define orchestration as ``the automated arrangement and coordination of complex networking systems, resources and services. It has an inherent intelligence and implicitly autonomic control of all systems, resources and services." They have a visionary aspect in their survey in terms of proposing multi-domain orchestration architecture, but they only focus on the components for realising multi-domain orchestration. They do not consider AI solutions for the end-to-end orchestration, and they do not include an edge computing aspect (storage and computing elements are seen as cloud nodes).

de Sousa et al. \cite{de2019network} provide a comprehensive survey of network service orchestration (NSO). They define NSO ``as the automated management and control processes involved in end-to-end services deployment and operations performed mainly by telecommunication operators and service providers, involving different types of resources and potentially multiple operators". In their definition, orchestration is responsible for decoupling high-level service/application layer from the underlying physical and virtual resources, and orchestration is divided into three functionalities: service, resource and lifecycle orchestration. Their taxonomy for NSO solutions has seven main categories: 
service models, software, resource, technology, scope, architecture, and Standards Development Organization (standardization activities in scope of NSO). Their view is limited on service life cycle orchestration in cloud platforms.

Zhong et al. \cite{Zhong2021} provide an extensive survey of ML-based container orchestration approaches. They form a taxonomy for ML-based container orchestration solutions. The taxonomy has five main categories: application architecture (the composition of containerized applications), cloud infrastructure, optimization objectives, behavior modelling and prediction (workload characterization, performance analysis, anomaly detection and dependency analysis), and resource provisioning operations for containerized applications. Their view is limited on the life cycle management of containers.

Versluis and Iosup \cite{Versluis2021} provide a systematic survey on workflow orchestration, focusing on taxonomizing four areas: workflow formalism, workflow allocation, resource provisioning, and applications and services. Formalism is concerned with the language used to represent workflows. Allocation places the workflows onto available resources in such a way that the scheduling targets are met. They see that allocation has five main categories: scheduling target, optimization strategy, scheduler structure, allocation technique and workflow instantiation. Resource provisioning is concerned about making decision when to allocate resources and how much given current and predicted demand. Applications and services are focused on the services provided by cloud, resource and environment types, and service execution models. Their paper has a strong focus on workload management in cloud infrastructures.

Costa et al. \cite{costa2022orchestration} systematically review fog orchestration literature and propose a generic architecture for fog orchestration. They define fog orchestration as ``a management function responsible for service life cycle. To provide requested services to the user and assure the Service Level Agreements (SLAs), it must monitor the underlying infrastructure, react timely to its changes and comply with privacy and security rules". They analyze 50 papers focusing on five aspects: the goals of orchestration, orchestrated entities, control topology, architecture layers, and orchestration functionalities. Based on the results, they formulate a generic fog orchestration architecture in terms of functionalities that an orchestration framework should fulfill. These functionalities are: 
admission control of incoming requests; service lifecycle management; resource management; resource monitoring; optimization to satisfy different objectives and constraints; communication management; a node agent to manage the node locally; and finally, security management. However, because their architecture is composed of the approaches found in the literature, they do not provide any concrete enablers towards realising all the components of the architecture.

\section{Definitions and Scope}\label{sec:Definitions}
\subsection{Artificial Intelligence and Autonomy}
AI applications are commonly defined as computational agents, that is, processes with autonomy and intelligence. Agents are capable of pursuing their objectives rationally, relying on their individual capabilities. They acquire information on their environment through perceptions (e.g. sensor data or messages) and, based on their autonomous decision-making process, can affect the environment with actions \cite{Mammela2018}.

An AI agent may possess various degrees of intelligence, defined by its reactivity, sociality, proactivity, and learning capability \cite{Weiss2013}. Reactivity refers to the agent's capability to react to changes in its environment, sociality to its interactions with other agents, and proactivity to its capability of taking initiative. Furthermore, learning allows the agent to evaluate its actions in the environment, and derive and explore new actions with the aim to reach its objective.

Agents possess individual knowledge of their (perhaps partially) observable environment, as well as of themselves and others. This knowledge is necessary for making decisions on actions and receiving and evaluating feedback of the actions, and subsequently, complex cognitive capabilities such as context-awareness and adaptivity in a dynamic open environment. In the scope of this article, such knowledge is encapsulated into objects referred to as models. The process of building and adapting these models, based on the agent's experience, is referred to as learning (or training).

Decision algorithms, which employ the learned models, encapsulate an agent's intelligence. Decision making uses the models to map objectives, percepts and previous actions to a new action which best increases the agent's probability of success. A popular choice is Reinforcement Learning (RL), which assumes the environment rewards for successful actions. RL models usually comprise one or both of an actor (or policy) model, which aims to find the appropriate action, and a critic (or value) model, which estimates the accumulating reward, given some choice of actions \cite{Grondman2012}.

Finally, autonomy here refers to an agent's ability 
to achieve its goal without any external control, by, e.g., users, administrators or other agents. In more detail, an autonomous system defines its own plan or direction for reaching the goal, making independent decisions along the way and readjusting its plan in case of obstacles \cite{Mammela2018}.

\subsection{Distributed AI}
Multi-agent systems (MASs) distribute the functionality of an application among a number of agents. Such distribution is necessary in an environment where individual agents do not have enough information or resources to achieve their objectives. Instead, the agents must cooperate on their individuals objectives and collaborate on shared ones, communicating their understanding of the environment and their progress towards the objectives.

Distributed application logic is implemented with agents that have distinct roles and behavior. The agents negotiate and share information and/or resources with each other to coordinate their efforts. Furthermore, a MAS can have an open or a closed organizational structure, which governs relationships, rules, objectives, policies and authority \cite{Weiss2013}. Such policies may comprise, for example, how agent interactions are conducted, and what an agent can expect from others. The behavior of a MAS thus emerges through the actions and interactions of autonomous or partially autonomous individual agents, with the guidance of an orchestrator or through a choreography of the autonomous participants. 

On open systems, such as the Internet of Things (IoT), a MAS often needs to dynamically reorganize itself to adapt and to evolve, in response to changes in the participating agents or in the environment. Individual and collaborative learning from past experience help MASs better to adapt and evolve towards proactive behavior in reaching common objectives.

\subsection{Computing continuum}\label{sec:ContinuumDefinitions}
\begin{figure}
    \centering
    \includegraphics[width=\linewidth]{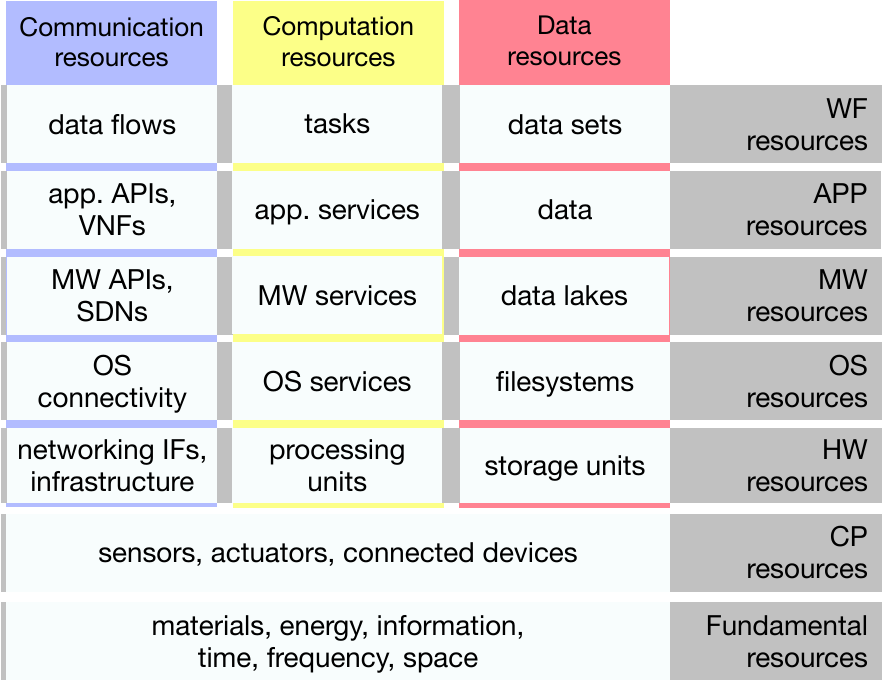}
    \caption{\textbf{Resources in the computing continuum.} The resources reside on any of a number of abstraction layers, from fundamental resources such as energy or time, to cyber-physical (CP), hardware (HW), operating system (OS), middleware (MW) and application layers (APP) or, finally, the layer of resources related to a particular application workflow (WF). In this hierarchy, higher-level resources rely on the lower-level ones, using them to fulfill their function. Further, the resources can be grouped as per their usage in three categories, namely, communication, computation and data resources.}
    \label{fig:orchestration-objects}
\end{figure}

Computational resources span over the network infrastructure, all the way from a central data center to the user at the edge of the network. Several architectural approaches, including fog and edge computing, Mobile Cloud Computing (MCC), and MEC \cite{Zhao2019, ranaweera2021survey}, take advantage of these resources, expanding the cloud computing paradigm. Although each approach encompasses its own paradigm and requirements, they bring services closer to the user while simultaneously addressing challenges inherent in edge application deployments such as latency requirements, bandwidth constraints or energy utilization.

The cloud, with ample resources for computing and storage, is still often a necessity, calling for hybrid edge--cloud architectures \cite{Xiong2018,Prabakaran2020,Yuan2020}, or even a continuum of computational resources between the devices and the cloud where applications can choose the best resource usage policy based on current needs \cite{Balouek2019continuum,Dustdar2022}. Accordingly, in this article, we collectively refer to edge and fog computing, MEC, and other similar distributed computing approaches with heterogeneous and opportunistic resources by the term \textit{computing continuum}.

Synthesizing the taxonomies in a number of recent works (see e.g. \cite{mampage2021holistic,costa2022orchestration, hong2019resource,tocze2018taxonomy,Zhong2021}, and those mentioned in \cref{sec:RelWork_Orch}), in this article, we take a holistic approach to the resources in the computing continuum. These resources, as depicted in \cref{fig:orchestration-objects}, are present on a number of levels, ranging from fundamental resources such as energy or time (see~\cite{Mammela2018}), through cyber-physical (CP), hardware (HW), operating system (OS), middleware (MW) and application (APP) resources, finally to workflow (WF) resources catering to the highest level application business logic or clients. In this hierarchy of levels, higher level resources rely on the lower ones to fulfill their function. Further, from hardware level up, the resources can be divided in three distinct categories, namely, communication, computation, and data-related.

It should be noted that this hierarchy of levels does not constitute a layered architecture in the sense that a level would only be aware of its immediate lower level. For example, data sets on the workflow level may be sourced from sensors on the cyber-physical level.

This holistic viewpoint is not present in the related studies. These studies (see \cref{sec:RelWork_Orch}) often refer to entities on hardware and operating systems levels as resources, and entities on middleware and application levels as services. However, there is considerable ambiguity in these conventions, and we find that an explicit consideration of EaaR (Everything as a Resource) simplifies the overall view.

\begin{figure*}
    \centering
    \includegraphics[width=\linewidth]{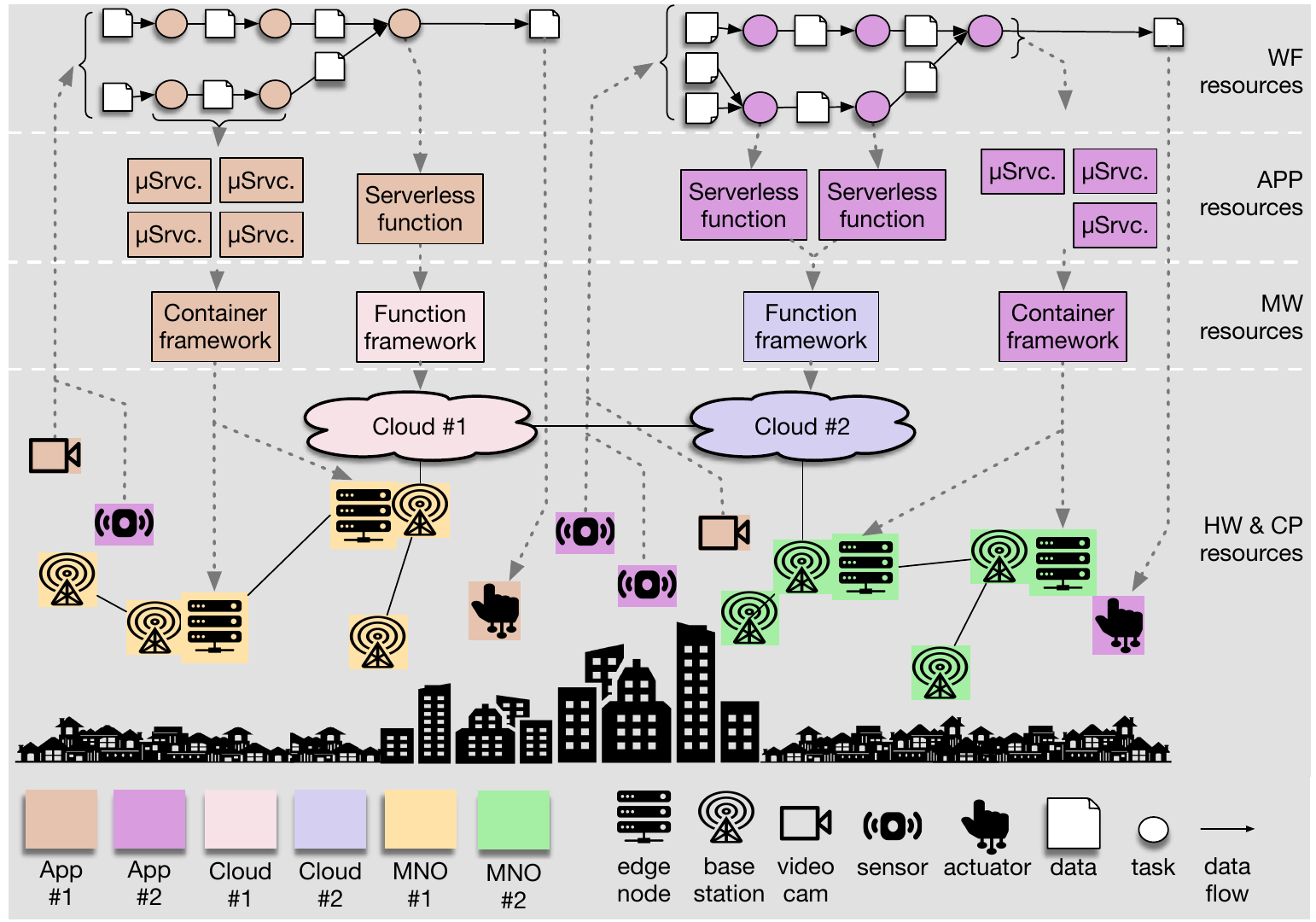}
    \caption{\textbf{Two example IoT applications in the computing continuum.} The workflows of two IoT applications, comprising tasks, data flows and data sets, are executed in containerized microservices and serverless functions, deployed on edge and cloud resources provided by Mobile Network Operators (MNOs) and cloud providers. Hardware (HW) and cyber-physical (CP) resources such as sensors, actuators, edge nodes and base stations are deployed in the vicinity of the users. Color coding corresponds to administrative domains.}
    \label{fig:computing-continuum}
\end{figure*}
    
The cyber-physical resources of a computing continuum may comprise sensors, actuators, and other connected user devices which have a physical form and function. They may act as sources (sensors) or sinks (actuators) of data flows in the computing continuum, or harvest fundamental resources such as energy from the surrounding environment \cite{zhang2021faster}.
    
Hardware resources in the communication category comprise, for example, physical network interfaces, access points, and base stations. The computational hardware resources refer to processing units (e.g., CPU, GPU, AI related accelerators), whereas data-related hardware resources include, for example, hard drives and SSDs.
    
Operating system resources include, for example, OS network interfaces and related abstractions such as sockets, operating system services such as processes (threads), and filesystems, as well as support for VM and containers. 

Middleware resources in the communication category include, for example, software defined networks (SDNs) and network slices \cite{Taleb2017,Afolabi2018}, and middleware APIs. Computational resources include middleware services offered to applications, such as those available in the edge/fog computing frameworks (e.g. ETSI MEC \cite{ETSIMEC2019}), as well as container frameworks such as Kubernetes or Docker Swarm (see \cref{sec:ContainerOrchestration}). Data resources in the middleware layer refers to, for example, databases, data warehouses, or data lakes, and distributed file systems. The middleware resources are provided through OS services, libraries, and local and remote APIs. 
    
Application resources include application network overlays, virtual network functions (VNFs) and application APIs. Moreover, application resources include services, often packaged as containerized microservices or serverless functions, and the data generated and consumed by the applications. 
    
Workflow resources are the highest-level category, consisting of the data flows, tasks, and data sets available for individual application workflows initiated by application clients or business logic.

A simplified example of two IoT applications in the computing continuum is depicted in \cref{fig:computing-continuum}. The workflows of the applications start with data lifted from sensors. The data is processed in a sequence of tasks, running on containerized services in edge devices or cloud-based serverless functions. The serverless function frameworks are provided by cloud service providers, while the edge nodes and communication infrastuctures are served by mobile network operators. Both workflows finally end on actuators. In the depicted example, the applications share some of their sensors with each other.

\subsection{Orchestration}\label{sec:OrchestrationDefinitions}

Managing the resources described in \cref{sec:ContinuumDefinitions} in the computing continuum is often referred to as \textit{orchestration}. In more detail, the term orchestration is used to refer to functions such as the automated management (i.e., configuring and coordination) of complex services, dynamic resource allocation, efficient and optimized resources utilization, control of functions, or real-time service delivery \cite{guerzoni2017analysis, Taleb2017}. However, related work often scopes orchestration to certain aspects of the continuum, such as networks and connections, application services, or tasks and workflows. Network orchestration thus refers to the configuration and management of communication networks. In contrast, service orchestration refers to the management and configuration of the life cycle of application components encapsulated as services.

\begin{figure*}[ht!]
    \centering
    \includegraphics[width=0.6\linewidth]{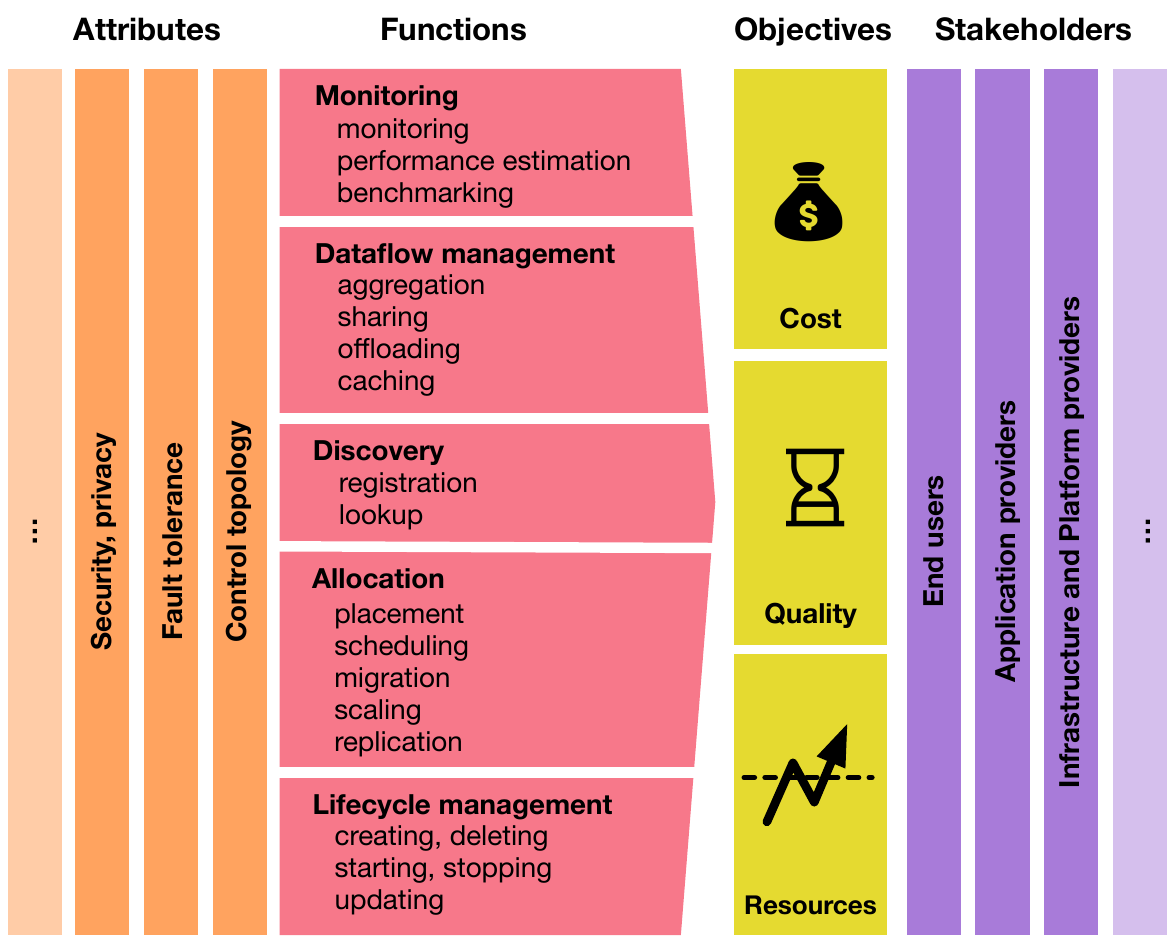}
    \caption{\textbf{Continuum orchestration.} Orchestration, that is, managing resources in the computing continuum, can be divided to a number of functions such as lifecycle management or monitoring, as well as overarching attributes such as security and privacy or fault tolerance. Orchestration aims to reach certain objectives, set by a number of possible stakeholders such as end users and infrastructure providers.}
    \label{fig:orchestration-taxonomy}
\end{figure*}

In this article, again synthesizing the concepts presented in recent studies on orchestration \cite{mampage2021holistic,costa2022orchestration, hong2019resource,tocze2018taxonomy,Zhong2021}, we holistically define orchestration as the management of resources in the computing continuum, from fundamental resources to workflow resources, as depicted in \cref{fig:orchestration-objects}. This management can be further divided into a number of distinct \textit{functions} (\cref{fig:orchestration-taxonomy}), with particular \textit{attributes}, aiming for a set of possible \textit{objectives}, set by different \textit{stakeholders}. 

Orchestration functions include:
    lifecycle management, comprising functionality such as creating, deleting, starting, stopping, or updating resources;
    allocation, comprising for example the placement of resources, scheduling access to them, or their migration, scaling and replication;
    discovery, supporting the registration and subsequent lookup of available resources;
    dataflow management, with functions to, e.g., aggregation, sharing, offloading and caching of data resources; and
    monitoring, which keeps track of the state and capabilities of each resource, and  estimates their performance.

These functions may be implemented in various ways. The implementations may differ from each other, and these differences may be described by attributes such as security, privacy, or fault tolerance. Moreover, orchestration may rely on control topologies including centralized, decentralized, and fully distributed \cite{costa2022orchestration}. The control topologies are discussed in more detail in \cref{sec:controltopologies}.

Dustdar et al. \cite{Dustdar2022} use Resources, Cost, and Quality as the three fundamental objective categories, which we adopt here. Further, the objectives of orchestration are often related to the essential resources (i.e., spectrum, time, energy, or funds) and can often be described as a combination of corresponding budgets or as a relation between the budgets and resource usage~\cite{tocze2018taxonomy}. For example, KPIs related to the QoS, most often expressed as latency or throughput, are related to the time budget for an application running in the continuum.

Finally, orchestration may be multi-domain, with a number of administrative domains, and many possible stakeholders setting the objectives. These stakeholders may comprise, e.g., end-users, owners of devices, application providers, or infrastructure providers such as mobile network operators (MNOs), Internet service providers (ISPs), or edge operators.

\subsection{Orchestration control topologies}\label{sec:controltopologies}

The three main control topologies of the computing continuum orchestration are centralized, decentralized, and distributed. The degree of the centralization of the control topology is intimately connected with the autonomy of the individual components in the system: a centralized control topology endows little autonomy to the local components, with a central controller making decisions, whereas decentralized and distributed topologies endow the components with increasing autonomy.

Centralized topologies are similar to classical master--worker models, defining an explicit and unique controller component that manages resource allocation and task assignment between the nodes. Liu et al. \cite{Liu2018} designed a centralized edge orchestrator able to overcome the resource limitations of individual mobile devices (nodes) by offloading computational tasks to an external server. The orchestrator receives task allocation petitions, and an optimization algorithm is invoked for optimal resource allocation in the worker external server. 

Analogously, Xiong et al. \cite{Xiong2018} presented KubeEdge as a hybrid cloud--edge platform where a centralized edge orchestrator was integrated for resource allocation from a cloud environment. The infrastructure was able to coordinate computing resources from the cloud to fulfill the requirements of the edge tasks, resulting in an improved performance in containerized applications.\footnote{While the KubeEdge approach is mostly centralized, some elements were related to a decentralized architecture, see discussion in \cref{sec:fogorchestration}.}

Conversely, in distributed architectures, nodes decide by themselves what tasks to allocate and how global resources are distributed based on the priority of individual tasks and resource availability.
Distributed control\footnote{Distributed orchestration is sometimes also referred to as \textit{choreography}, see e.g. \cite{peltz2003web,blanc2021service,al2022ai}.} is based on resource allocation algorithms that omit centralized edge controllers entirely. These algorithms are distributed over the nodes, which negotiate to reach a consensus on the optimal task allocation. Agreed tasks are then distributed over the computational resources available in the continuum.

A paradigmatic example is the algorithm developed by Castellano et al., called \emph{Distributed resource assignment and orchestration algorithm} (DRAGON) \cite{Castellano2019}. This resource allocation algorithm allows a set of applications to reach an agreement on how edge infrastructure resources have to be assigned without the need for a centralized orchestrator. Each application has a DRAGON agent that starts a voting procedure to acquire its resources to deploy a given task. Then, each application participates in a resource election protocol where the winning applications allocate their demanded resources on a certain physical node.

Another example of distributed control is the resource distribution algorithm developed by Fizza et al. \cite{Fizza2018}, which allocates incoming tasks by prioritizing deadline requirements while taking into account privacy constraints. The Privacy Aware Scheduling in a Heterogeneous Fog Environment (PASHE) algorithm can schedule resource allocation into heterogeneous edge devices (each with a different computation capability) by offloading public loose-deadline requiring tasks into centralized cloud environments and private urgent tasks to be performed locally.

Other aspects apart from resource capabilities can be considered during orchestration or even prioritized over resource management by distributed control topologies. An example of this is the work by Auluck et al. \cite{Auluck2019}, who developed an orchestrator that dynamically allocated the incoming tasks attending to deadline and security aspects. Workloads are performed in edge data centers (more insecure) or cloud data centers (more secure) depending on the security tags of each task and the expected delay of the network. 

Decentralized approaches are in the middle of centralized and distributed, where clusters of hierarchically organized nodes manage the computing continuum resources. Intra-cluster a centralized edge controller orchestrates the resources, while inter-cluster a distributed, peer-to-peer, approach is used \cite{costa2022orchestration}.

An example of this is the work by Tocze et al. \cite{Tocze2019}. They designed an edge orchestrator for task allocation (ORCH Framework), which addresses the problem of Distributed Dynamic Task and Mobile Edge Placement (D2TEP). The ORCH framework divides the Edge area into edge device neighborhoods, each orchestrated by an area orchestrator. By combining stationary and mobile edge devices, the framework can flexibly serve all edge tasks (either mobile or stationary).

\subsection{Scope}
This survey is located in the intersection of distributed AI and computing in the device-edge-cloud continuum. We focus on decentralized and distributed control topologies for the orchestration of the computing continuum, modelling a deployment in the continuum as a network of connected resources, on a number of abstraction levels as depicted in  \cref{fig:orchestration-objects,fig:orchestration-taxonomy}. 

In more detail, orchestration in the computing continuum is increasingly complex. To address this complexity, and to increase fault tolerance, a system in the continuum must adopt autonomy. In other words, the local components of the system must be able to operate autonomously, not relying on resources at other nodes, while still keeping in synchronization with their global objectives.

We thus model the computing continuum system as consisting of autonomous, intelligent agents. These agents may reside on any of the computing tiers and resource abstraction layers (\cref{fig:orchestration-objects}), and they act as service endpoints in their own roles, as advocated by Xiong et al. in their vision of decentralized internet of things~\cite{Xiong2018}. Each of these agents maintains its objectives, and acts autonomously to maximize its selfish interest, i.e., to meet its objectives as best as it can. 

In short, we consider systems in the computing continuum to comprise a MAS, and be composed of intelligent and autonomous or semi-autonomous agents with high reactivity, proactivity, sociality, and learning capabilities.

Deep learning can be regarded as a significant key factor behind smarter edge platform management \cite{Wang2020}. Hence, this survey mainly assumes that the agent models are Artificial Neural Network (ANN) models.

Finally, this survey also touches upon the security and privacy aspects in developing an autonomous computing continuum. The main emphasis will be on what security benefits and issues the integration of AI with the computing continuum environment causes.

\section{Orchestration Paradigms}\label{sec:Edgeorchestration}
This chapter provides an overview on the orchestration mechanisms relevant to the computing continuum. The taxonomy followed from \cref{sec:ContainerOrchestration} to \cref{sec:networkorchestration} is in terms of the computing resources, while \cref{sec:fogorchestration} and \cref{sec:emergingparadigms} are devoted to solutions specifically targeting the Edge and Fog paradigms. In brief, each section presents a general description of the main concepts, followed by state of the art examples and finished with a discussion of their suitability for the computing continuum.

\subsection{Container orchestration}\label{sec:ContainerOrchestration}
A container is a standard unit of software that encapsulates the processes and dependencies of an application service. 
Containers allow running microservice based applications reliably, in isolation of the processes of the underlying operating system or other containers, and enable easy migration from one computing environment to another \cite{khan2017key}. Furthermore, containerization can be deployed to provide computational capabilities for systems in the computing continuum, enabling batch processing at scale, control planes, and IoT and AI workloads \cite{casalicchio2019container}. As such, container orchestration focuses on the application and middleware resource layers in \cref{fig:orchestration-objects}.

Microservice-based applications that often run in clusters of hundreds of geographically distributed containers must be fault-tolerant and 
easily accessible \cite{khan2017key}. Managing the life-cycle of such a high number of containers, scaling them up and down, as well as replicating, migrating, starting, and stopping them as necessary is 
a complex task \cite{khan2017key}. To overcome this complexity, container orchestration systems have been developed to manage the lifecycle and workflow of the containers in an automated manner in large and dynamic environments \cite{al2019container}. 

In more detail, container orchestration refers to controlling and automating the deployment of containerized resources 
considering different requirements such as availability, scaling, and networking. The functions related to container orchestration include scaling containers up and down across the host infrastructure, scheduling and managing clusters, migrating containers from one host to another, allocating resources between the containers, load balancing, monitoring containers, managing networking overlays, and providing security \cite{khan2017key}. 

In addition, container orchestration must meet certain standards 
such as secure networking 
without impacting the local network, and low requirements on resources such as memory, processing power, and storage \cite{goethals2019fledge}. 
Thus, to meet these requirements, container orchestration minimizes the use of communication, computation, and storage resources such as memory, CPU/GPU, disk space, volumes (i.e., interacting with local and remote file systems), as well as ports and IPs that refer to the configuration of IP addresses and application ports in containers. While allocating and guaranteeing the necessary resources for containers, container orchestration must also optimize the resource usage requirement of the orchestration process itself. 

Currently, there are several orchestration platforms available in the market, with the open-source Kubernetes\footnote{https://kubernetes.io/}, Docker Swarm\footnote{https://docs.docker.com/engine/swarm/} and Apache Mesos\footnote{https://mesos.apache.org/} the most widely-used. 

In more detail, Kubernetes is an open-source container orchestration platform introduced by Google to automate, deploy, manage and scale containerized applications across a cluster of nodes\footnote{https://delftswa.gitbooks.io/desosa2018/content/kubernetes/chapter.html}. 
Kubernetes is based on a centralized control topology, with a master-slave architecture. The Kubernetes master node defines a control plane, managing the cluster by deploying applications to slave nodes, while the slave nodes host and run the application containers. To operate, Kubernetes uses a set of objects called Pods, which are the basic control unit of Kubernetes. Pods consist of containers that share the 
processes, interfaces, IP addresses, ports, and memory. 

Moreover, Kubernetes manages resource scheduling and networking. Resource scheduling enables detecting and utilizing the available pods at the nodes, and networking facilitates communication between containers (i.e., pods) using container network interfaces through overlay networks \cite{kayal2020kubernetes}. 

Docker Swarm is a container orchestration framework built especially for Docker\footnote{https://www.docker.com/}-based containers. Docker Swarm functions by clustering and scheduling Docker containers deployed across multiple host nodes. 
Docker Swarm shines in its fast deployment, simplicity, and high availability \cite{cerin2017new}. 

As with Kubernetes, the control topology of Docker Swarm is also centralized, with a the master-slave architecture. 
To schedule a container for a slave node, Docker swarm takes into account the resource requirements (such as CPU and memory) of the service it will deliver. Scheduling strategies include the default \emph{Spread} strategy, which selects the node having the least number of containers, the \emph{Binpack} strategy, that selects the most packed containers, and the \emph{Random} strategy that randomly selects a slave \cite{cerin2017new}.

Docker swarm also employs an overlay network to connect containers that are hosted on different nodes. The network allows the master node to determine which slave nodes are still functional. Furthermore, Docker Swarm has load balancers on every node, balancing workload across nodes and containers.

Apache Mesos, originally developed by the University of California, Berkeley, 
manages application services in large-scale clustered environments. Apache Mesos brings the resources of different nodes into a single pool of resources, eliminating the need for assigning specific nodes per workload. To do this, Mesos abstracts the hardware resources such as CPU, memory, and disk space available on the nodes, and manages access to these resources to prevent processes from interfering with each other \cite{frampton2018apache}. 

In summary, Kubernetes, Docker Swarm, and Apache Mesos all rely on a centralized control topology. These orchestration platforms are examples of the cloud-native container orchestration paradigm, emphasizing container life cycle management while mostly disregarding the continuum environment characterized by, for example, intermittent communication, heterogeneous and geographically distributed computing, and siloed and non-IID data. According to Costa et al.~\cite{costa2022orchestration} (among others), these limitations make them and, indeed, the current container orchestration paradigm as a whole unsuitable for taking full responsibility for the orchestration in the computing continuum. However, a container orchestration system could be used as a component of a broader orchestration framework in the continuum.

\subsection{Workflow management systems}
Workflow management systems refer to software systems that facilitate the coordination of workflows. In more detail, workflow management systems offer an infrastructure for setting up, specifying, executing, and monitoring workflows, which comprise tasks that run on available computing resources  \cite{casati1999specification,yu2005taxonomy,Mitchell2019}. 


Based on Versluis and Iosup \cite{Versluis2021}, workflow management comprises a formalism, that is, a language such as CWL, DAX, or YAWL to describe workflows, as well as systems implementing the formalism. Furthermore, a workflow management system implements a number of different functions such as scheduling, optimization, and instantiation of workflows and tasks. Workflow management may have different targets such as cost, latency, resource usage, or load balancing.
As such, workflow management systems follow closely the framework described in \cref{sec:ContinuumDefinitions}. The difference comes in the scope, i.e., workflow management focuses on the workflow resources and application resources layers in \cref{fig:orchestration-objects}, while largely omitting those below.

Apache Airflow\footnote{https://airflow.apache.org/} is an Apache Software Foundation project developed by Airbnb in 2016. Apache Airflow is designed to offer a lightweight workflow management system to model, maintain, and monitor workflows. Airflow executes each workflow as a directed acyclic graph of tasks. These tasks are usually atomic, run independently, and do not exchange data with each other. Airflow can scale horizontally on clusters orchestrated by Apache Mesos. It can be utilized to execute workflows in diverse computing platforms such as workstations, edge, and cloud platforms. It also offers multiple interfaces to commonly used cloud environments such as Amazon S3, Google Cloud, or HDFS. This feature enables the users to access and utilize multiple clouds only with one deployment \cite{bernstein2014containers}.

Kubeflow\footnote{https://www.kubeflow.org/} is an open-source workflow orchestration tool based on the Kubernetes container orchestration environment. Kubeflow is designed to easily deploy ML workflows on  Kubernetes clusters. Kubeflow functionality echoes that of Kubernetes, as it aims to deploy, scale, and manage ML workloads \cite{george2022end}. 
Kubeflow also allows running automated ML tasks and supports hyperparameter tuning \cite{george2020scalable}, thus supporting end-to-end ML workflows \cite{zhou2019katib}. 
In addition, Kubeflow aims to evaluate the end-to-end performance of deployed workflows in terms of resource consumption through GPU utilization and time requirements for CI/CD (i.e., continuous integration and continuous delivery) pipelines \cite{zhou2020towards}.

MLflow\footnote{https://mlflow.org/} is another open-source workflow management platform, developed by Databricks. 
MLflow is designed to manage and streamline the complete ML lifecycle, enabling users to bring their software and workflows into the ML lifecycle \cite{chen2020developments}. 
MLflow is composed of four major components, namely, Tracking, Projects, Models, and Model Registry. These components can be utilized individually or together. The Tracking component is an API that records experiments, and it is used for tracking data such as parameters or input data. MLflow Models is a generic model packaging format that can operate across diverse ML environments. MLflow Projects is a packaging format that uses a YAML configuration file for packaging software into reusable projects. MLflow Model Registry component is a collaborative hub that manages the lifecycles of ML model deployments \cite{zaharia2018accelerating}. 

However, none of the above workflow management systems consider the communication, computation, and data-related challenges in the computing continuum. 
In literature, there is only one proposal related to learning on heterogeneous devices, on non-IID data~\cite{Lo2021}. However, that proposal is only restricted to FL systems.

\subsection{Network orchestration}\label{sec:networkorchestration}
Network orchestration refers to the automated management and control of the deployment and operation of end-to-end services in telecommunication networks~\cite{de2019network}. In more detail, network orchestration focuses on the communication resources resources column \cref{fig:orchestration-objects}, managing for example SDNs, VNFs, and network slices, which control low-level resources (such as bandwidth and networking interfaces) available on the fundamental, hardware, and OS resource layers~\cite{Galis2009management, de2019network}.  

In essence, network orchestration maps user service requests to underlying resources on a number of layers, thus scheduling access and sharing the virtual and physical resources in the network \cite{etsi2014network}. Further, network orchestration is responsible for managing the life cycle of network services such as VNFs, using a set of orchestration functions such as registering, instantiating, scaling, updating, and terminating~\cite{etsi2014network}.

Currently, open standardized network orchestration is an emerging area for research. In literature, many studies are devoted to efficient resource utilization and orchestration in networks.
For example, Hirwe et al.~\cite{hirwe2016lightchain} aim at minimizing load balancing using a run time procedure in order to optimize VNF placement and service chaining in NFV. 
Kuo et al.~\cite{kuo2018deploying} aim to minimize resource utilization through optimizing a joint problem of VNF placement and routing in the network. 
Pham et al.~\cite{pham2019multi} propose a game-theory-based theoretical model to capture the competition among network service providers in a multi-domain NFV. 
Nejabati et al.~\cite{nejabati2021zero} introduce a zero-touch network orchestrator that automatically orchestrates and manages network resources and generates performance profiles of the network services.

Furthermore, Salhab et al.~\cite{salhab20215g} propose a predictive approach for network slicing, aiming at optimal decisions for network slicing to dynamically utilize shared network resources.  
Bari et al.~\cite{bari2016orchestrating} identify the optimal number and placement of VNFs to improve operational costs and network resource utilization. To orchestrate the virtual resources and network functions, Guerzoni et al.~\cite{guerzoni2014novel} propose embedding a virtual network to deal with the mapping of virtual resources on the physical infrastructure in a network. Finally, Khan et al.~\cite{khan2020generic} propose an intent-based system that automates the orchestration of end-to-end workflows and services across multiple orchestrates, by developing an algorithm based on graph neural networks that estimates resource requirements for VNFs. 

Indeed, all of the studies presented above focus on orchestrating the communication resources, and to some extent also the computational and data resources used by  communication-related services (e.g., VNFs), with an aim of keeping network service level objectives (SLOs). However, in these studies, there is little consideration towards considering the resources on the application and workflow layers,  
that need to balance their communication requirements against their computational and data related ones. 
Moreover, even within the orchestration of communication resources, interoperability between vendors and operators remains an open issue~\cite{rotsos2017network}. Finally, the underlying SDN solutions are predominantly centralized~\cite{benzekki2016software,de2019network}.

\subsection{Edge/fog orchestration}\label{sec:fogorchestration}
The edge/fog orchestration concept aims to address the shortcomings of the above paradigms in the computing continuum. As such, edge/fog orchestration emphasizes the role of the cyber-physical resources in conjunction with the communication, computation and data-related ones, and stresses their scarcity in the computing continuum (see \cref{fig:orchestration-objects}). 

For example, Costa et al.~\cite{costa2022orchestration} draft a common architecture for fog orchestration, synthesizing dozens of proposals. The draft emphasizes access control, service and resource management, monitoring, optimization, and communication management. While the draft covers  most of the functionalities presented in \cref{fig:orchestration-taxonomy}, it does not consider container life-cycle management, or the orchestration of the lower layers of network resources. 
Further, at this early stage, the proposed architecture is still missing the necessary details such as its control topology.

In addition to academic proposals (see, e.g., \cite{tocze2018taxonomy,hong2019resource} in addition to \cite{costa2022orchestration}), a number of frameworks are already available for developers or exist in various stages of standardization. For example, the ETSI Multi-access Edge Computing (MEC) standard \cite{ETSIMEC2019,sabella2019developing}  defines an orchestration architecture for Multi-Access Edge Computing. In the ETSI MEC architecture, a centralized orchestrator is responsible for monitoring, lifecycle management, scheduling, and migration. Further, this standard defines interoperation with, e.g., cloud systems, defining functionality such as migration of applications between cloud and MEC \cite{ETSIMEC2021}. However, ETSI MEC is centralized by design. Further, originating from the telecom industry and focuses on mobile networks, it may struggle with the orchestration of the fog domain or low-resource IoT devices.

Moreover, a number of approaches such as MicroK8s\footnote{https://microk8s.io/}, KubeEdge\footnote{https://kubeedge.io/en/}, and OpenYurt\footnote{https://openyurt.io/} extend the container orchestration framework Kubernetes towards edge devices, albeit with different architectural approaches. MicroK8s provides a lightweight but full Kubernetes implementation, extending the cloud to the edge with enough resources to support the streamlined framework. KubeEdge and OpenYurt both introduce a new abstraction layer for the edge resources and connect this layer to a cloud-based main Kubernetes cluster. However, while KubeEdge implements the edge layer with custom components, OpenYurt instead adopts Kubernetes native plug-in and operator mechanisms for the implementation.

Unlike MicroK8s, which relies on Kubernetes' centralized control topology, with their newly-introduced edge layers, both KubeEdge and OpenYurt are somewhat akin to a decentralized one. Indeed, edge clusters in KubeEdge and OpenYurt have limited autonomy in case their cloud connections are severed. However, both architectures rely on containerization, which is often unsupported on the most lightweight devices. Including these devices in the orchestration framework would thus require custom solutions.

\subsection{Emerging paradigms}\label{sec:emergingparadigms}

\subsubsection{Multi-domain orchestration}
Multi-domain orchestration is an emerging computing paradigm that aims to provision end-to-end network service delivery across multiple infrastructure providers, each one having their own administrative domains. The paradigm is particularly useful for upcoming resource-hungry applications, such as holographic applications or tactile internet, where provisioning necessary computing, storage, and communication resources that meet the required quality of service by the application developers and the desired quality of experience by the end users may not be located within an administrative boundary of a single cloud service provider. As such, multi-domain orchestration studies the effects of multiple stakeholders for the attributes, functions and objectives of orchestration (see \cref{fig:orchestration-taxonomy}).

Current standard architectural Frameworks, such as ETSI Management and Orchestration (MANO) \cite{craik2019network} provide tools and setups to provision VNFs, their configurations, and deployment within the infrastructure of a single cloud service provider. The ETSI MEC architecture \cite{ETSIMEC2019,sabella2019developing}, discussed earlier in \cref{sec:fogorchestration}, allows applications to be virtualized at the edge and access the network. This virtualization at the edge within ETSI MEC framework could be implemented as VNFs within the ETSI MANO framework. However, both of these architectures fall short in assembling functionalities provided by multiple service providers into one single function which could operate over multiple infrastructure in end-to-end fashion, owned by different service provider. 

Recent work by Francescon et. al. \cite{francescon2017x}, X–MANO presents a proof-of-concept for cross-domain network service orchestration. It introduces an information model which allows network network service provider to advertise its resources to other network service providers in a privacy preserving manner. For this, it introduces a Multi–Domain Network Service Descriptor (MDNS) which lets service provider to expose its network services without revealing internal implementation details. It introduces programmable network service to enable network service providers to implement their custom life-cycle management polices.

Genez et. al. \cite{6676686} proposes scheduling mechanism based on time discretization with controlled granularity to allocate resources for a service from using resources owned by multiple service providers. It models resource scheduling as a directed acycle graph where nodes represent specific jobs of the service and edges represent dependency among them, and provides a solution for the scenario where these nodes may be owned by different administrative domains.

Osmani et. al. \cite{9627828} proposes an extension to Kubernetes, a popular choice for container orchestration. It proposes a Federated Kubernetes framework for building a unified management mechanism for multiple Kubernetes clusters, where each individual Kubernetes cluster is owned by a different service provider, and is subject to their control polices. To build this unified management mechanism, it proposes the Network Service Mesh, a framework that providers a secondary network connectivity among Kubernetes containers, and services various roles such as client, endpoint, control manager, data manager in different domains to facilitate inter-domain interactions. Moreover, Subramanya and Riggio \cite{subramanya2021centralized} propose AI-driven Kubernetes-based orchestration which uses time-series forecasting techniques to manage dynamic resource up-scaling in multi-domain scenarios.

From a resource orchestration point of view, multi-domain orchestration is an ideal option for resource-hungry applications in the upcoming metaverse age, where resources available locally may not be enough. However, it poses challenges in terms of privacy, security, heterogeneity, and in aligning conflicting interests of different service providers. While a fully-fledged multi-domain orchestration system could be the final realization of the computing continuum system, the current efforts focus mostly on either network or container orchestration. These cover only a subset of the full range of orchestration requirements; further, containerization is often not supported on the wide range of lightweight devices without custom solutions. 

\subsubsection{FaaS orchestration} 
Function as a Service (FaaS) is a cloud computing service under the emerging paradigm of serverless architecture. FaaS allows application developers to develop, run and manage individual, isolated application functionalities at the cloud~\cite{Barcelona-Pons2019}. It offloads the responsibility of building and managing the infrastructure associated with developing and running the application from the application developer to the cloud service provider. Developers may write and update application code on the fly as microservices, which can then be executed in response to events. Moreover, these events may be triggered by the application user in real-time, with, say, user clicking a particular element in the application GUI. As such, FaaS emphasizes the workflow and application resource layers in \cref{fig:orchestration-objects}.

In a nutshell, FaaS enables application developers to focus only on developing the application, while the cloud service provider takes the complete responsibility of the resource allocation and run-time management, as well as the security measures. As such, FaaS offers the benefits of faster development turnaround to the developer. Furthermore, FaaS promises built-in scalability, as the cloud service provider manages application resources, and scales them up under sudden bursts of demand. In addition to simplifying developing, this also reduces costs \cite{schleier2021serverless}. 

These benefits make FaaS ideal for applications, especially those which provide basic utilities, such as Google Maps. However, FaaS has its disadvantages. Application developers may need to write their applications to be compatible with the back-end infrastructure of the cloud service provider, enforcing a vendor lock-in. Moreover, debugging and testing become complex since developers always need to cope with changes in the infrastructure of the cloud service providers.

Over recent years, many commercial platforms and open source platforms supporting FaaS have emerged. Among commercial platforms, Amazon AWS Lambda holds the leading position in terms of market share and a range of services it enables. In FaaS, application developers define a workflow which may consist of a single or multiple functions \cite{mampage2021holistic}. This workflow can be interpreted as a state machine. To orchestrate the state machine, cloud server uses lightweight services, encapsulated as containers or micro-virtual machines, which execute the functions defined in the state machine when triggered by an event, say, a user clicking on an application component which invokes the function. Each service may host one or more functions depending on the amount of memory each function requires. Orchestration of the services follows the strategy of bin packing problem, placing different functions to services in order to maximize memory utilization. While FaaS platforms follow this strategy on a high level, they differ in their implementation details, such as their strategies for resource allocation, billing, and tools provided to application developers to write the workflow of their functions \cite{lopez2018comparison}.

From an orchestration point of view, FaaS partly simplifies  resource management, due to the statelessness of the services, while also extending the scope of container orchestration towards that of workflow management. Moreover, in the computing continuum, the lightweight FaaS functions can be hosted at nearby IoT or edge nodes, thus benefiting objectives related to, e.g., latency. Statelessness could enable different functions of any given application to be hosted on devices across different tiers, thus offering the benefit of portability. 

However, FaaS poses a challenge for applications with heavy context or state data, such as Augmented Reality (AR)/Virtual Reality (VR), or those employing AI/ML models. Indeed, migrating these functions may need moving large data sets across the network \cite{lee2021all}. Furthermore, orchestrating workflows over these functions requires careful consideration of how state information in the workflows is maintained. While the full view of FaaS orchestration is still largely an unsolved problem, developers and cloud service providers may need to consider caching the data geographically close their user base, a strategy which is often used by video streaming applications like Netflix \cite{deng2017internet}.

\subsubsection{WASM}

WebAssembly (WASM) is a portable low-level bytecode which offers a compact representation, efficient validation and compilation, and safe execution with little overhead \cite{Haas2017}. While originally designed to run in web browsers, WASM has recently gained traction as a general-purpose compilation target for a number of programming languages and run environments \cite{Makitalo2021}.  

As such, WASM could provide an alternative to containerization, especially in resource poor edge devices incapable of supporting containers or container orchestration, and especially for FaaS workloads \cite{Gadepalli2020}. 
While some early cloud based orchestration frameworks for WASM exist (see, e.g., discussion on WASMcloud by Rac and Brorsson \cite{rac2021edge}), to the best of our knowledge, there are currently no orchestration frameworks available for applications based on distributed WASM runtimes in the computing continuum.

\section{Research Themes and Challenges}\label{sec:Vision}
In this section we look at certain emerging research themes that may affect the future of orchestration, and synthesize the themes to provide an early vision of a future continuum orchestration paradigm. Furthermore, we describe the challenges related to the emerging research themes and the vision.

\subsection{Computing Continuum management model}\label{sec:vision_mgmt}

Computing continuum invalidates the idea of having single-tier, centralized orchestration, such as exclusively employed on the cloud. On the contrary, the challenges in computing continuum, discussed in \cref{sec:synthesis}, push towards distributing orchestration along the tiers, addressing management from a holistic perspective~\cite{Dustdar2022}. 

Indeed, the complexity and scale of the computing continuum challenge current methodologies for managing distributed Internet-based systems. Unlike Cloud systems, the continuum cannot be considered as \textit{elastic}~\cite{dustdar2011principles}, rendering online Cloud-Fog-Edge computing as an outdated management methodology. General and adaptive modelling of the Computing Continuum Management (CCM) is thus in a key role in building novel solutions.

In particular, each stakeholder in the computing continuum sets its own objectives for the operation of applications. However, it remains an open question how these objectives and their fulfillment can be measured or enforced. Furthermore, the computing continuum comprises a high number of resources on many different computing tiers and layers of abstraction. Maintaining the objectives while keeping the system in balance is also an open question.

In more detail, Morichetta et al.~\cite{Morichetta2021} recognize for example the following particular research topics in relation to CCM:

\textit{Flexible representation.} To orchestrate the resources in the computing continuum, CCM needs a flexible and adaptive representation of those resources. Since the architecture of the system may change, with new resources appearing and disappearing opportunistically, the representation must be able to reflect these changes.

\textit{Operational link to an underlying infrastructure.} Any application running in the computing continuum is highly dependant on the underlying resources. Since those resources may be heterogeneous and dynamic, comprising for example a wide variety of different IoT devices as well as Edge, Fog, and Cloud configurations, any CCM methodology has to consider the infrastructure as a key component.

\textit{Temporal evolution.} The computing continuum is constantly changing and evolving. Any management model must consider this change, and allow for concept and data drift. Moreover, the rate and direction of this change (say, in terms of the orchestration objectives related to costs, quality and resource usage) may also be considered to allow for appropriate reporting and consequent action.

\textit{Causality relations.} The computing continuum comprises an ecosystem of multiple interacting resources and stakeholders. A global perspective thus has to consider the whole of this ecosystem, not restricting to individual resources or their activities. To understand how actions propagate across this ecosystem, CCM needs to keep track of causal relationships between the resources.

\textit{Proactive adaptation.} The complex causal relations between the resources may lead to a cascade of failures as issues can propagate across the computing continuum. Maintaining stakeholder objectives thus requires prompt and proactive action to prevent such failure propagation. 

\textit{Emergence.} Causality and complexity, when not properly managed, can together cause harmful emergence and interference between system components. These issues are discussed in more detail in the following subsection \cref{sec:WeakCoupling}.

\textit{Learning framework.} The ecosystem complexity and scale make it impossible to draw a complete management plan in the design phase. Therefore, setting management methodologies inside a learning framework is required to provide incrementally better solutions and adaptations. 

\subsection{Weakly coupled, autonomous control}\label{sec:WeakCoupling}
A novel orchestration paradigm must strike a balance between local autonomy and centralized control~\cite{Mammela2022}. Indeed, local autonomy brings all the benefits discussed above but it is not sufficient alone. This is because achieving system goals is a known challenge when agents are completely autonomous or even if they cooperate locally with each other. The main reason is emergence: due to nonlinear interactions, the system level behavior cannot be predicted from the behavior of the individual agents. Autonomous continuum agents might not get the resources they need, or the agents might interfere with each other. The goals might not be reached at all, or even if reached, resource usage might be too high~\cite{riekki2021research}. 

Hence, some centralized control is needed to reach system goals with sufficient performance. In other words, loose (weak) coupling is required: the agents are nearly autonomous, but fair resource allocation and agent cooperation are ensured via minimal centralized control. Such minimal centralized control enables achieving system goals that require compromises from the agents. Furthermore, centralized control allows global optimization: balancing resource usage so that the goals are reached and performance requirements fulfilled~\cite{Mämmelä2021}.

Loose coupling leads to a hierarchical system where the higher levels control the lower ones, operating at lower resolutions and having wider perspective to the system and the environment~\cite{Mammela2022}. The highest levels can be located in the cloud. The higher levels can use Multi-Objective Optimization (MOO) to realize fair resource allocation for the lower-level agents, i.e., a Pareto-optimal allocation. This optimization can be learnt over time and updated/re-learnt when changes in the system or its environment invalidate what has been learnt. MOO can help to reach goals while fulfilling constraints on resource usage. Loose coupling is a favorable property also for interaction between agents at the same hierarchy level -- the interaction should be minimized.

Loose coupling introduces the same benefits for both applications and edge orchestration. Multi-agent systems are a natural choice for loose coupling. The centralized control can be realized through goals and constraints (specifically on resource usage). The agents have otherwise local autonomy but they advance the goals given by the higher level and do not violate the given constraints. The higher levels monitor the progress and update the goal and constraints when needed, based on the optimization they perform.

From the perspective of continuum orchestration, the open questions in loose coupling are related to the implementation of orchestration functionality with autonomous agents. The general question is: How can the targeted system behavior be achieved with a set of independent, loosely-coupled agents? More specifically, the following topics need to be studied, among others:

\textit{Degree of centralization.} Due to its benefits, local decision-making should be favored and centralized control used only when necessary. What is the optimal balance between independent local decision-making and centralized control in a large system of systems? How does this balance vary over time, for example, when abrupt changes are experienced in the operation environment?

\textit{Emergence.} In a distributed, loosely coupled system, patterns of activity between the agents may emerge. How can harmful emergence be avoided in such systems? How can emergence be used in achieving the system-level goals?

\subsection{Semantic communication in the computing continuum}\label{sec:vision_semantic}

Current communication paradigm employs information theory to quantify the maximum data rate that can be supported by a communication channel. This theory was developed in the 1940s by Claude Shannon \cite{shannon1948}, and it has been guiding the design of the information systems up until 5G and strikingly continues to be so even in the current mushrooming 6G visions.

However, this traditional focus has completely disregarded the semantic aspects of communication, viewing the meaning of the messages as largely irrelevant to communication \cite{Lan2021semantic}. In particular, as distributed intelligent agents interact, new notions of information are required to describe the common context between the agents and the potential of the agents to learn to efficiently communicate \cite{Foerster2016rial} changes in that context. 

One possibility is to base this notion of information on the von Neumann--Morgenstern theory of sequential games~\cite{von1944theory}, where information refers to imperfect knowledge of state. In more detail, as distributed agents interact, a shared context emerges, resulting in the exchange of information states that are minimally sufficient for the accomplishment of common goals. This would reduce the amount of information in the classical sense that agents would need to transmit or receive. However, the prerequisites for such efficient joint behavior emerging are not known, and neither is the mathematical characterization of such networked systems. 

Other fundamental problems related to semantic communication include the following topics, among others: 

\textit{Emergence.} How and under what conditions cooperative communication among agents emerges and is robust to deviations between agents having different priors/beliefs/knowledge/architectures? 

\textit{Convergence.} Under which conditions will agents converge to a shared language when learning from data? 

\textit{Scarcity.} How to deal with the fact that knowledge will be limited per agent in terms of incomplete information, limited compute power, memory for training, inference, reasoning, planning?

\subsection{Security and Privacy}\label{sec:vision_security}
Edge AI, that is, distributed AI methods deployed in the computing continuum, can play a vital role in enhancing the security and privacy of future orchestration in the computing continuum. Added intelligence in the continuum allows the deployment of advanced and smart security mechanisms at the edge of the network, closer to the external users and devices. Thus, overall security can be improved as it is possible to detect and mitigate some attacks via external user devices already at the edge of the network \cite{zhang2018data}. The early detection of some attacks will be essential to eliminate the impact and the propagation of such attacks to the critical core network elements.

Moreover, edge AI based methods are processed in local vicinity of the user instead of in the remote cloud, reducing the need of transferring raw user data between the user and the Cloud via untrusted backhaul networks such as the Internet~\cite{ranaweera2021mec}. Moreover, local processing eliminates the possibility of various security attacks such as Sybil, Denial of Service (DoS)/Distributed Denial of Service (DDoS), fiber tapping, hidden pulse attacks, jamming attacks, as well as privacy leakages associated with a long-distance data transmission phase~\cite{ranaweera2021survey}. 

The heterogeneous ecosystem of IoT is suffering from various security issues due to the lack of device processing capabilities, improper security standardization, energy-saving strategies, lack of expertise, and untrusted device manufacturers~\cite{xiao2018iot}. Thus, the implementation of advanced security mechanism at the network-level as Security-as-a-Service (SECaaS)~\cite{boudi2019assessing, 9148660} is one of the viable solutions to address the above limitations. Edge AI methods can implement a more advanced and intelligent SECaaS mechanism for IoT devices~\cite{alsharif2021study}. Added intelligence along the computing continuum allows deploying advanced security mechanisms such as threat intelligence tools, intelligent deep packet inspectors, and firewalls~\cite{haddadpajouh2020ai4safe}. For example, for limiting the effects of DDoS attacks, edge AI methods can provide a network perimeter around cloud resources. This perimeter could then intelligently adapt the packet processing rate at the router to stop the cloud from being overwhelmed with DDoS requests, time-out half-open connections, and equip the router with an intelligent filter to drop the packets that are likely to be part of the attack.

Furthermore, edge AI can be used to implement AI algorithms in a decentralized manner. Thus, some attacks on edge AI systems can be restricted to the local vicinity, not impacting other edge devices. In this way, edge AI can also eliminate the possibility of a single point of failure associated with centralized AI algorithms~\cite{verma2018distributed}.

Finally, managing and processing user data locally increases~\cite{ding2021roadmap} or even guarantees user privacy, given the applied distributed learning and decision making processes employ privacy-guaranteeing approaches such as differential privacy~\cite{porambage2021roadmap}.

Some particular research topics related to security and privacy issues are listed below:

\textit{Physical security.} One of main challenges is the physical security of the AI components. AI integration with edge computing will promote edge AI resources to a critical role in network management and orchestration~\cite{mukherjee2020intelligent, yahuza2020systematic}. However, deployments in the computing continuum typically reside outside the central data infrastructure, in physically insecure premises in the wild. The computing continuum should thus have extra precautions in place to mitigate the impact of physically tampering with devices, adding malware to edge devices, swapping or interchanging devices, or injecting rogue data sets and AI algorithms. Additional security measures such as hardware root of trust, crypto-based ID, encryption for in-flight and automated patching should be deployed to obtain  tamper-proof edge AI devices and services. As an example, Sachdev et al.~\cite{sachdev2020towards} discuss the critical security and privacy issues for edge AI in IoT/IoE digital marketing environments along with some possible mitigation mechanisms.

In addition, the impact of capturing edge devices with AI is much more severe than capturing an edge device without AI due to added intelligence and distributed control functionalities. If an attacker takes control of an edge AI device, he might be able to jeopardize or take control of almost all the localized network services of a particular local vicinity. Moreover, decentralization of edge AI opens up unknown security issues such as synchronization attacks, information protection issues, membership inference, and API-based attacks~\cite{verma2018distributed}.

\textit{Trust.} As various stakeholders can connect their resources to the computing continuum~\cite{du2018big}, establishing trust between the stakeholders can prove challenging. Ensuring trust in multi-domain orchestration is thus an important factor for beyond 5G networks. 

Ensuring the trust in the edge AI system is challenging due to its distributed nature and the potential liability issues~\cite{ding2021roadmap}. Strategies such as open implementation, well-defined specifications, and reputation systems can be used to address these trust issues~\cite{marcus2019rebooting}. In addition, it is necessary to define and differentiate the control role between devices by the user and resources provided by providers (e.g., AWS, GCP, Azure). Further, the improved privacy benefit of keeping the user data local at the edge servers is only achieved if the edge servers have higher privacy and trust levels than the cloud server. Since the edge by itself does not guarantee privacy, this might raise new challenges to ensure the privacy of user data~\cite{ding2021roadmap}.

\textit{AI confidentiality and integrity.} Integration of AI/ML techniques will lead to a new set of AI/ML related attacks such as data injection, data manipulation, logic corruption, poisoning, evasion, model inversion and membership inference attacks~\cite{bertino2021ai, siriwardhana2021ai}, which edge AI systems need to mitigate, calling for novel defence methods~\cite{porambage2019sec}. 

For example, to prevent the leakage of personally identifiable information, the training data needs to be anonymized or protected with cryptographic techniques. In addition, the final machine learning model needs to be protected against model inversion and membership inference attacks. Anynomization techniques, such as differential privacy, guarantee that personal information does not leak from the training data or from the final model. However, such anynomization methods can lead to a reduction in classification performance, as well as a lower convergence rate~\cite{Wei_2020}. They also have limitations when applied on unstructured data, such as images~\cite{Liu_2021}. Cryptographic techniques do not have such limitations, but incur a high computational or communication cost. Research is needed to determine the optimal tradeoff between privacy protection and the performance of the system.

As another example, the training of accurate models requires that the training data is also accurate. Poisoning attacks influence the training data resulting in generalization errors in the final classifier. While there are methods that attempt to detect and exclude poisoned samples for simple models, such as linear regression~\cite{Jagielski_2018}, the poisoning problem is still largely unsolved~\cite{Pitropakis_2019}. In particular, for FL, poisoning attacks can lead to a substantial decrease in classification accuracy and recall even with a small number of poisoned model updates~\cite{Tolpegin_2020}. Novel methods are needed to detect poisoned samples and local model updates before the global model computation, as well as to prevent misclassification through evasion attacks.

Another integrity issue arises from the model training process. In edge computing, training is potentially distributed or executed on an untrusted entity. Malicious parties need to be prevented from inserting false training data or a backdoor into the final model. Homomorphic encryption~\cite{Rivest_1978} can be applied to train the model on encrypted data and verifiable computing~\cite{Gennaro_2017} can attest its correctness, but both require significant computational resources from the computing party or the edge devices. An alternative approach is secure multiparty computing~\cite{Yao_1986}. However, due to its interactivity it incurs a significant communication cost and typically does not scale well with the number of participants~\cite{Saia_2015}. The tradeoff between computing and communication costs needs to be carefully considered especially when both computing power and bandwidth are limited.

\subsection{Synthesis}\label{sec:synthesis}
Synthesizing the above research topics, we can provide an early vision of decentralized, multi-domain, multi-tenant orchestration in the computing continuum. In our vision, resources in the continuum are modelled as agents, and each agent is trying to fulfill externally set objectives on cost, quality and resource usage. While trying to reach those objectives, the agents make decisions on when and how to conduct actions related to orchestration functions (see \cref{sec:OrchestrationDefinitions}). Moreover, to conduct those actions, the agents may need to negotiate with other agents to provision resources. As an example, an agent may need to make a decision on offloading computations to cloud from an edge node, and needs to negotiate the terms of the offloading with the cloud agent to understand the consequences of that decision.

\begin{figure}[b!]
    \centering
    \includegraphics[width=0.6\linewidth]{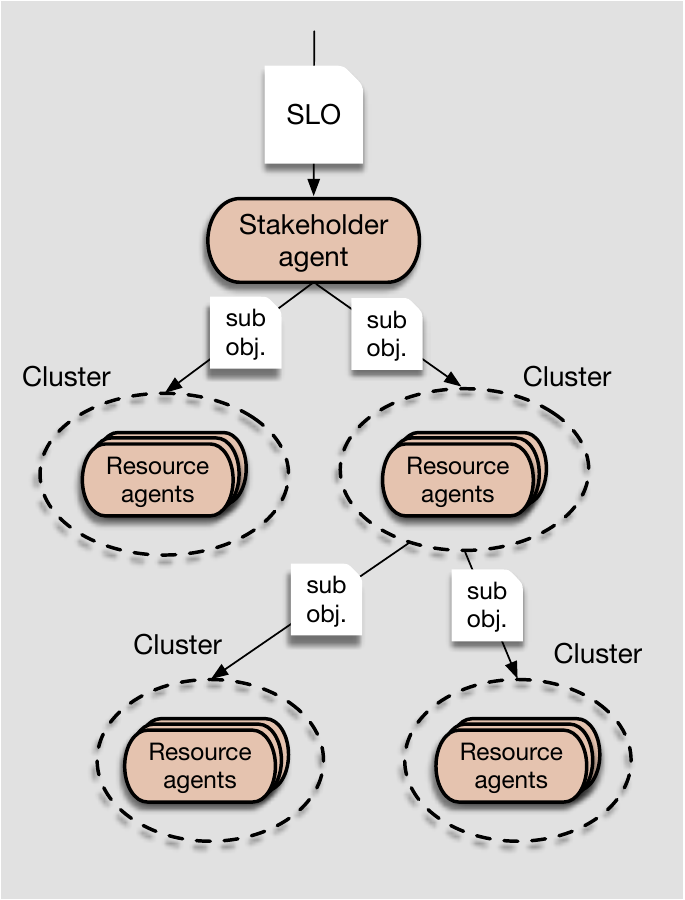}
    \caption{An example of the objective overlay of decentralized orchestration. The application stakeholder agent pursues its objectives, related to cost, quality and resource usage, breaking them down for sub-objectives for high-level clusters. These clusters, in turn, oversee the agents in their sub-clusters.}\label{fig:vision-objective}
\end{figure}

The objectives are set on the agents according to their place in a hierarchy. Each stakeholder, be that cloud or network service provider, or an app, is represented by a stakeholder agent on the highest level of the hierarchy. These agents have service level objectives to fulfill, related to cost, quality, and resource usage, set by the application developer or domain manager. The stakeholder agents break down and pass on their objectives to the agents below them in their administrative domain, organized as clusters and further sub-clusters (\cref{fig:vision-objective}).

The distance measure for forming these clusters may vary based on the environment. For example, forming clusters based on proximity minimizes latency and maximises connectivity. Furthermore, the clusters need to be dynamic to handle the opportunistic nature of the continuum. As an example, the highest level cluster agents could be cloud and edge cluster agents, with edge sub-clusters comprising individual locations of interest, with new agents appearing and disappearing as new devices move through that location.

Meanwhile, agents negotiate for resource usage within as well as across administrative boundaries. Within administrative boundaries, the negotiation can be considered co-operative, while over the boundaries negotiation may be competitive. While negotiating, the resource agents need to consider their own objectives, broken down from those of their stakeholder agent and passed to them along the control hierarchy (\cref{fig:vision-objective}).

This framework follows the principle of \textit{weak coupling}, with a hierarchy of agents. Each agent makes autonomous decisions on how to reach its objectives (\cref{sec:WeakCoupling}). These objectives are constructed from those of the stakeholders in the computing continuum (\cref{sec:vision_mgmt}). Over administrative boundaries, the agents may need to \textit{learn to communicate} for their negotiation (\cref{sec:vision_semantic}). Studying further,  the \textit{security} of the architecture and the  \textit{privacy} of the users must be considered from the very beginning (\cref{sec:vision_security}).

The computing continuum poses a number of challenges in realizing the vision of orchestration based on autonomous, intelligent AI agents. Communication is intermittent and fluctuating, computation resources are distributed, heterogeneous and opportunistic, and data is distributed, siloed and non-IID, and at times sensitive. Finally, there can be a massive number of resources, applications, tenants and other stakeholders over a number of domains, and these stakeholders may have partially conflicting objectives. 

\begin{figure}[b]
    \centering
    \includegraphics[width=\linewidth]{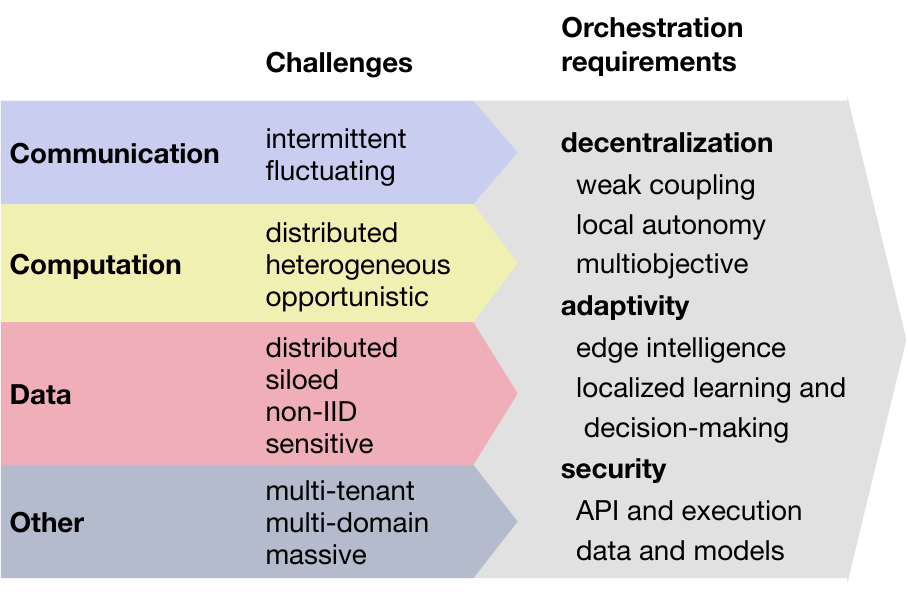}
    \caption{Challenges inherent in the computing continuum, and subsequent characteristics required of the AI approaches.}\label{fig:challenges_requirements}
\end{figure}

These challenges set requirements for the orchestration of the computing continuum, as depicted in \cref{fig:challenges_requirements}. Orchestration must be decentralized, as the resources are; further, weak coupling \cite{Mämmelä2021} and local autonomy allow the approaches to survive alone if connections are severed. Non-IID data requires distributed edge intelligence, with localized learning and decision-making, while the numerous stakeholders and tenants present in the continuum demand approaches that support balancing multiple objectives. Finally, security and privacy must be considered for APIs and execution, as well as for data and AI models.

However, implementing AI methods to answer these requirements is challenging. In particular, the following topics must be considered:

\textit{Common context and standards.} While AI agents may learn to communicate and negotiate, they need some common context and standards to ensure interoperability and security of transactions. For example, distributed learning of models requires that AI agents share information on their models with each other. What is the minimal common context (e.g., protocol for sharing model information, or means of describing model metadata such as structure or purpose) required for this sharing of model information? How is resource discovery implemented? Can learning be fully distributed, or is a central authority required to fulfill some function?

Moreover, for multi-agent and distributed decision making, an agent may need to know what the other agents observe, how they act, and how in general they make decisions. The extent to which this is possible or reasonable may vary due to, for example, the availability of computational capacity, data, or communication links. Is it possible to define a minimal set of resources required for decision making? Can the methods be scaled based on available resources?

Finally, what is the minimum amount of common knowledge agents can be assumed to have at the beginning of negotiation? What type of common ground can be instantly established for negotiation within and across administrative boundaries?

\textit{Convergence.} The computing continuum is a challenging environment for conducting distributed learning. Is it always possible to guarantee convergence in model learning, or do the agents have to accept that in some cases, learning does not result in a usable model? What consequences does this have for the agent's subsequent decision-making? Moreover, what are the conditions under which one can guarantee a convergence into a stable and optimal decision making strategy, considering also the possible adaptation of the strategy when the underlying environment changes?

\textit{Lack of training data.} Data in the computing continuum is often geographically distributed and locked behind siloes. Furthermore, online training of a decision making strategy from scratch may be unfeasible. The algorithms are data hungry, and can make very awful decisions particularly during initial learning stages.
Building training data sets in such an environment may be challenging, calling for pre-training, or for methods that use unlabeled data, or learn quickly from few samples. What is the best approach in each case?

\textit{Adaptivity.} The computing continuum is in a constant state of flux, with nodes appearing and disappearing, connectivity fluctuating, users moving, and application components migrating between nodes. AI methods need to be able to scale their resource usage to adapt to this change. However, on the other hand, AI methods must strike a balance between stability and adaptation, not reacting to temporary changes. What is required to identify the balance point here?

\textit{Security.} Edge AI methods may alleviate security concerns in the computing continuum, bringing advanced capabilities close to the proximity of users. However, AI methods may also introduce novel vulnerabilities. Does this cat-and-mouse game converge towards an acceptable equilibrium? What novel security architectures are needed in the computing continuum to ensure data and model integrity throughout their lifecycles?

\section{Survey of Edge AI Approaches}\label{sec:Enablers}
This section presents promising AI approaches which address some of the challenges described in \cref{sec:synthesis} and may lead into solutions that consider all of them. In particular, we look at distributed and secure methods for ML, decision making, and negotiation, and finally consider shortly certain other emerging approaches. We will be focusing more on introducing research fields that we believe will be the key areas of the research for the future computing continuum orchestration rather than on providing an in-depth analysis of the methods inside the fields due to the enormous amount of work each field entails.

\subsection{Distributed Learning}\label{sec:distlearning}
Orchestration in the computing continuum can be modelled as a hierarchical network of intelligent, autonomous agents that manages the resources of the platform in a decentralized manner. These agents need ML models to make predictions about processes and future states, which supports the agents' decision making processes. Based on data, the agents must learn different dynamics in their environment to improve their performance and to adapt to the uncertain, evolving environment.

Each agent has access to the data they have collected, but this data may not have enough volume or diversity to train accurate models. In addition, an edge agent may not have enough resources for the training and inference of complex models. Hence, it is inevitable that agents must somehow collaborate with other nearby agents in the training and inference of ML models. 

Below, we introduce methods that, with further development, can be key enablers for achieving efficient distributed and decentralized ML model training and inference among a network of AI nodes, addressing some of the challenges described in \cref{sec:synthesis}. The focus is on supervised learning methods.

\begin{figure*}
    \centering
    \includegraphics[width=\linewidth]{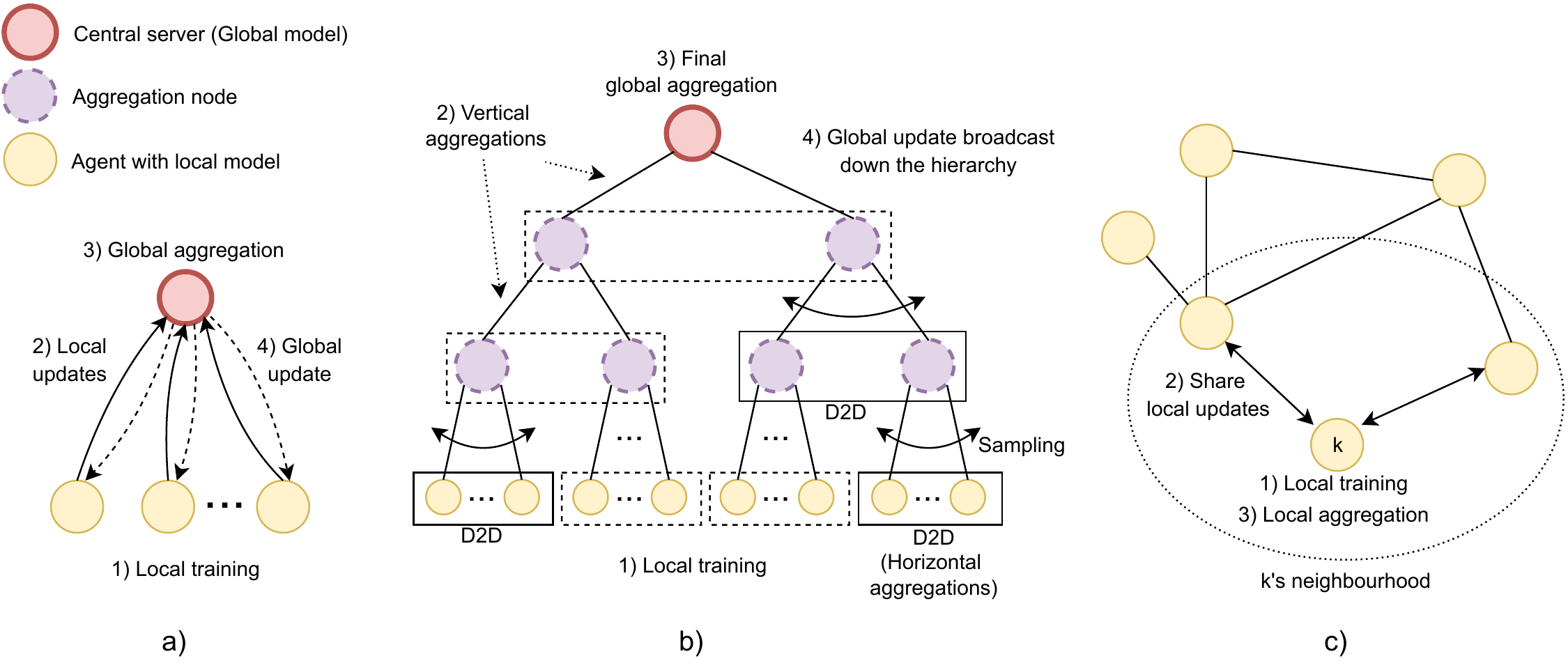}
    \caption{\textbf{High-level workflows and topologies.} \textbf{a) Federated Learning:} The nodes form a star topology where agents with local models periodically send their updates to a central server that aggregates them and broadcasts the global update back to the nodes. Same topology applies to Federated Distillation, where instead of model updates model outputs are exchanged. \textbf{b) Fog Learning:} The nodes form a cluster-based hierarchy, where aggregations travel through layers to the central server. D2D enabled clusters aggregate the model updates among themselves, after which the cluster's parent fetches the aggregated result form one of the nodes. Non-D2D enabled clusters perform aggregation as in FL. \textbf{c) Peer-to-peer Learning:} Generally, the nodes can form an arbitrary network graph, where each node exchanges their local updates in their neighbourhood. The workflow in this figure is a high-level description of ATC diffusion.}
    \label{fig:dist_learning}
\end{figure*}

\subsubsection{Federated Learning}
Training ML models on edge requires distributed learning architectures and algorithms. 
FL has quickly become the de facto training paradigm for distributed model training in edge environment. FL, first introduced by Google \cite{McMahan2017}, aims to train a global ML model in a distributed manner. The global model is most typically an ANN model, but it can also be some other parameterized model. Original version of FL, often called \emph{vanilla FL}, trains a global model in a centralized manner on decentralized data. Each agent participating in the training has their own training data that they use to train a local model. Then, the local parameter updates are sent periodically to a central server that aggregates the updates and sends the resulting global model back to agents (\cref{fig:dist_learning}a). 

In \emph{Federated Averaging} (FedAvg), the baseline FL algorithm introduced by McMahan et al. \cite{McMahan2017}, the global update is calculated as the weighted average of the local parameters. Each agent calculates the local parameters by applying Stochastic Gradient Descent (SGD) locally a specified number of times after the previous global parameter update. FedAvg can also be implemented based on exchanging gradients instead of the parameters, in which case the central aggregator calculates the mean of the local gradients that is subsequently used by the agents to update their parameters locally \cite{Park2019}.

Vanilla FL approach has multiple issues and deficiencies, which render it incompatible with our vision of an autonomous computing continuum. Starting from communication efficiency, wireless links with low bandwidth and intermittent connectivity are an essential part of the computing continuum environment, while communication is a major bottleneck in FL. The communication costs of vanilla FL can be partly controlled with the update period that determines how often local parameters are sent to the central aggregator. For example, Wang et al. \cite{wang2018} propose a control algorithm for dynamically changing the update period so that global loss function is minimized under a fixed computational and communicational resource budget. However, exchanging model parameters (or gradients) can cause huge communication overhead, as modern ANN models can have millions of parameters. One common solution is to reduce the size of the transmitted update by model compression schemes, which include, for example, model sparsification or low-rank approximations, quantization of parameters, lossy compression, and Golomb lossless encoding \cite{Li2019}.

Agents participating in training are usually heterogeneous in terms of hardware, as well as resources such as energy or network connectivity. This can cause stragglers in the learning; agents may take too long to respond or completely drop out. As a consequence, some agents can have outdated models. Solutions to the straggler issue include asynchronous communication, 
actively selecting the participants to the training based on system resources, and increasing the fault tolerance of training to drop outs through algorithmic redundancy with methods such as coded computation \cite{Li2019}. Another solution is to offload the tasks and data to a neighbouring node even though this requires considerations of privacy and affects the learning latency, as well as the accuracy of the model \cite{Park2021}. On the other hand, if the dynamics of resource availability are modeled as a time series, Gaussian Process Regression (GPR) can be used to predict future resources, which will allow identifying agents that are more likely to be stragglers in advance \cite{Park2021}. Model parameter dynamics can also be predicted with GPR, which allows agents to estimate the model parameters of others in order to continue local training under erratic connectivity.

Incentive mechanisms are important for building accurate models with honest participation. Agents with higher quality data should be incentivized to participate in the learning, and they should also be rewarded for their greater contribution. In competitive settings, in particular, each agent gaining the same global model regardless of their contribution to the training is not desirable. The incentive can be, for example, a final model with a different level of performance, that is, the quality of the model is proportional to the level of contribution \cite{lyu2020}. Additionally, blockchain technology provides many secure incentive mechanisms \cite{Nguyen2021}.

When the goal is to train a global consensus model, non-IID data across the agents causes significant problems because distributed training algorithms usually assume that the data is IID, and convergence guarantees are given with the assumption of IID \cite{Li2019}. Data augmentation is one technique for mitigating the non-IID data problem. In its simplest form, data augmentation can be the sharing of data between the agents. For example, in the experimental study by Zhao et al., they were able to increase the model accuracy significantly by sharing only 5\% of the agents' local data \cite{zhao2018federated}. However, such sharing of data compromises privacy and communication efficiency, as well as undermines one of the key ideas of distributed learning, which is to keep the training data local. In an approach termed \emph{federated augmentation}, a server trains and shares a generative model that can be used by the agents to augment their data locally \cite{jeong2018}. 

To ensure the convergence under non-IID data and varying system resources, Li et al. propose an algorithm called FedProx \cite{li2020fedprox}. It introduces two modifications to FedAvg in order to achieve this. First, it adds so called proximal term to the local objective function, which addresses non-IID data by restricting the local model updates to be closer to the received global model at the beginning of a training iteration. Then, it introduces a notion of inexactness, which allows to incorporate also partial work (from stragglers) in the global update. In other words, a given number of local updates is specified to be carried out before global aggregation, but if an agent has not had enough resources to calculate all the updates, it can send its partial solution for global aggregation nevertheless. Li et al. show through their experiments with five different non-IID data sets that FedProx significantly improves the convergence rate when compared to FedAvg even when 90\% of the agents are stragglers.

How to train an accurate and reliable global model with low latency is a crucial question, especially in distributed learning on edge, where the training procedure encounters issues of non-IID data, energy and memory limitations, as well as low communication bandwidth. Park et al. provide an in-depth overview of different theoretical and technical enablers for low latency federated training and accurate inference under on-device and communication constraints \cite{Park2019}. They focus particularly on reliability guarantees, latency reduction and scalability enhancement in distributed model training. They present, among other things, suitable mathematical tools for quantifying generalization error bound of a federated model, e.g., with meta distribution, and for examining whether training loss reaches a target loss level, e.g., with extreme value theory.

\subsubsection{Federated Distillation}
Traditional distributed learning algorithms exchange model parameters. An approach termed as \emph{Federated Distillation} (FD) exchanges model outputs instead of model parameters, which reduces communication costs \cite{jeong2018}. FD has been developed for supervised classification. In FD, each agent keeps track of the local average logits per label, sending them periodically to a central server. The central server calculates the global average logits per label, which are then downloaded by the agents. Finally, each agent uses the global average logits in a distillation regularizer that is added to its local loss function. The regularizer penalizes larger differences between a sample's local logits and the global average logits of its true label. 

The communication payload size of FD is proportional to the output dimension rather than the model size, which reduces communication costs substantially. FD also allows each agent to have a different model, e.g., a agent with less resources can have a simpler neural network. However, FD is more vulnerable to non-IID data distribution across agents and produces less accurate models when compared to FL \cite{Seo2020}.

The communication-accuracy trade-off in FD, that is, even though the communication costs are substantially reduced, the accuracy is usually worse when exchanging model outputs instead of parameters, can be exploited in mobile communication systems, where uplink data rates are usually much lower than downlink data rates. The idea is to utilize FL in downlink (agents download model parameters) and FD in uplink communication (agents upload model outputs) \cite{Oh2020mix2fld}. The central server must do a model output-to-parameter conversion by utilizing knowledge distillation to update the global model it maintains. The server minimizes the difference between the uploaded outputs and the outputs of the global model, which naturally requires a set of samples. For this, agents also send a set of seed samples to the server, which have been encoded to preserve privacy.

\subsubsection{Fog Learning}
Hosseinalipour et al. introduce a paradigm called \emph{Fog Learning} \cite{Hosseinalipour2020, hosseinalipour2020MH-FL}, which shifts from the centralized star topology FL towards a semi-decentralized architecture. They develop a multi-layer cluster-based learning architecture especially suitable for collaborative model training across the computing continuum (\cref{fig:dist_learning}b). The nodes form a hierarchy, where the nodes holding the data and training the local models are at the bottom, the main server for global aggregation is at the top, and the model parameter aggregations and the parameter updates travel across multiple layers. The nodes in each layer form clusters, and inside each cluster collaboration can be enabled with device-to-device (D2D) communication. Nodes on the same level may dynamically change the clusters they belong to during the model training.

For training a global model inside the hierarchical, cluster-based network, Hosseinalipour et al. develop an algorithm called \emph{multi-stage hybrid federated learning} (MH-FL) \cite{hosseinalipour2020MH-FL}. As the main server is only interested in the weighted average of the local parameters, MH-FL is based on local aggregations at each network layer, which can happen either through distributed aggregation or instant aggregation. Distributed aggregation is for D2D enabled clusters, where all nodes can participate in a consensus scheme that ensures all devices inside the cluster agree on the average of their model parameters. The parent node of the cluster needs to then only fetch parameters from one node inside the cluster, and scale the obtained parameters by the number of nodes in the cluster to approximate the sum of the nodes' parameters. Instant aggregation is used in the case where D2D communication is not enabled inside a cluster. It is similar to the conventional aggregation of FL, that is, the parent node collects and aggregates the parameters from the nodes in the cluster.

The learning accuracy of MH-FL is able to approach that of the centralized gradient descent \cite{hosseinalipour2020MH-FL}. The experiments also show that MH-FL can result in 50\% device energy savings on average and 80\% reduction in the number of parameters transferred over the network layers compared with conventional FL.

\subsubsection{Peer-to-peer Learning}\label{sec:declearning}
FL relies on central aggregator, which creates a single point of failure, compromises scalability and maintains higher communication costs due to the star topology of the network. Our vision of an autonomous computing continuum requires less reliance on centralized entities and more independent and decentralized ways to coordinate model training.

In decentralized learning, agents are organized as a virtual and possibly dynamic network graph with a small maximum node degree, and the nodes can only share their local model parameters in their neighbourhood (\cref{fig:dist_learning}c). These local models should gradually converge to a global model, i.e., agents should reach \emph{consensus} \cite{kairouz2021advances, vanhaesebrouck2017}. The optimization problem is usually formulated as a global consensus problem \cite{Boyd2011dist_opt}, which allows to split the global model optimization into separate problems that can be solved in a distributed manner across agents. This formulation requires the addition of an equality constraint to ensure that the local parameters are equal. 

The global consensus problem can be solved with a primal-dual method called Alternating Direction Method of Multipliers (ADMM) \cite{Boyd2011dist_opt}. However, ADMM handles the equality constraint by relying on a central parameter server that has to collect all local parameters from the agents. To achieve truly decentralized model training, Elgabli et al. \cite{Elgabli2020} propose a primal-dual method called Group ADMM (GADMM), which is a communication-efficient decentralized algorithm for solving the global consensus problem. GADMM divides the agents into head and tail groups, where each agent in the head group communicates only with two agents in the tail group and vice versa (i.e., the overlay topology is a logical chain). Consequently, at each communication round, only half of the agents are competing for the limited communication bandwidth. Further, this topology allows to rewrite the equality constraint so that each agent only has joint constraints with two neighbours, except for the `end' agents that have only one. At each training iteration in GADMM, the head agents first update their primal variables (parameters) in parallel, after which they send the updated parameters to their tail neighbours. Then the tail agents update their primal variables in parallel, and the updated parameters are subsequently sent to the head neighbours. Finally, all the agents update their dual variables locally. 

Elgabli et al. \cite{Elgabli2020} also extend GADMM to Dynamic GADMM (D-GADMM), where the set of neighbours to each agent varies over time, meaning that the overlay topology is still a logical chain, but the physical neighbours can change. Hence, in D-GADMM, all the agents periodically check their connections, and if some sort of change is detected, they broadcast their parameters to the new neighbours. After this they find their new logical neighbours in the chain and send the right dual variable to their right neighbour to ensure that they share the same dual variable. D-GADMM can also be used to change the logical chain topology even when the physical topology itself does not change, which improves the convergence speed of GADMM while preserving the significant reduction in communication costs, as shown by Elgabli et al. \cite{Elgabli2020}.

Besides primal-dual methods, also primal methods have been proposed for solving the distributed optimization problem, even though these approaches formulate the consensus problem as an unconstrained problem that depends on a shared global variable. Primal methods often utilize Distributed Stochastic Gradient Descent (DSGD), and there exist different approaches based on how the model parameters are exchanged and aggregated between the peers. It is good to note that at least for static distributed optimization problems, where the local objective functions do not vary with time, primal-dual methods have been consistently outperforming primal methods \cite{Yuan2020dec_opt, Elgabli2020}.

In primal approaches that utilize gossip protocol, agents select one or a few of their neighbours for model parameter exchange in order to propagate model updates across the network. The choice of neighbours can be done randomly or in a more optimized manner with the utilization of virtual topologies. However, purely random gossip algorithms usually lead to convergence issues in large scale systems \cite{Deng2019}. To this end, Daily et al. propose GossipGraD \cite{daily2018}, an algorithm that scales DSGD on large scale systems by combining asynchronous communication with gossip partner selection and partner rotation to ensure faster propagation of model updates and faster convergence. GossipGraD was developed particularly for training DL models on distributed memory systems, which consist of a set of processing nodes interconnected by a high-speed network. Furthermore, as noted by Savazzi et al. \cite{Savazzi2020}, early approaches on gossip based methods, such as GossipGraD, cannot be fully used in D2D networks because they ignore medium access control and half-duplex constraints.

Diffusion is another primal approach for solving the distributed optimization problem. There are two main strategies for diffusion: Combine-Then-Adapt (CTA) and Adapt-Then-Combine (ATC) \cite[pp. 456-469]{sayed2014book}. In CTA diffusion, at every iteration, each agent first calculates the weighted average of the local parameters in its neighbourhood (including the agent itself), obtaining an intermediate result (\emph{combination step}). Then the agent uses this intermediate result to perform a gradient descent update (\emph{adaptation step}). In ATC diffusion, the combination and adaptation steps are reversed. Now at every iteration, each agent first uses the parameter estimate from the previous iteration to perform a gradient descent update, obtaining an intermediate result. Then the agent calculates the weighted average of all the intermediate results in its neighbourhood.

CTA diffusion propagates the information across the whole network more thoroughly, because the information is diffused by evaluating the gradient vector at the aggregate value of the neighbourhood's local parameters. On the other hand, ATC diffusion also includes information from the neighbours' data, because the combination step incorporates the gradient vectors of the neighbors.

More general diffusion approaches exchange gradients in addition to model parameters. The purpose is to increase the information flow and speed up the convergence by utilizing also neighbours' local data more efficiently, even though these approaches lead to increased communication costs \cite{Chen2012, Savazzi2021}. Here, rather than taking the gradient descent step based only on the gradient of the agent, we take the weighted average of the whole neighbourhood's gradients evaluated at the same point (aggregated value for CTA, local parameters from previous round for ATC). This naturally increases the cooperation and communication between the agents, because it adds computation and communication steps, where agents have to now also calculate their gradients at the aggregated value/local parameters of their neighbours and provide the results to them.

With the aim of reducing communication costs of gradient exchange, Savazzi et al. propose Consensus based Federated Averaging with Gradient Exchange (CFA-GE) \cite{Savazzi2020}, which combines elements from average consensus methods \cite{Olfati2007consensus} and CTA diffusion. CFA-GE has been especially developed for training ANN models in large scale networks with intermittent connectivity.

CFA-GE consists of four stages. At the beginning of each iteration, each agent receives the local model parameters of its neighbours from the previous iteration and combines them with its own using consensus-based aggregation. Each agent then sends the resulting intermediate result back to its neighbours, each of whom calculates a gradient evaluated at the intermediate result using the local data. After an agent has received all gradients from its neighbours, it calculates a weighted average of them and uses the average to update the intermediate result (one gradient descent step), obtaining an aggregated model. Finally, each agent does a number of SGD updates with its local mini-batches starting from the aggregated model, obtaining the final local model parameters to be sent to the neighbours for the next iteration.

The CFA-GE algorithm described above requires many communication rounds, as well as synchronization because each agent has to wait for the gradients from its neighbours. Hence, Savazzi et al. \cite{Savazzi2020} also propose a two-stage asynchronous algorithm for implementing CFA-GE. This improved algorithm is achieved by removing the stages of sending the intermediate results and then waiting for the neighbours' gradients. Now, these gradients are predicted using the past intermediate results from the neighbours, meaning that at the beginning of each iteration, each agent now receives the previous intermediate results and the gradient predictions from its neighbours. Each agent can use this information to finish the whole local model update without communicating with other agents, and at the end of the iteration it will simply send its new intermediate result, as well as the gradient predictions it made to its neighbours.

Utilizing distributed ledger technologies in combination with FL provides another way to decentralize the model training. Integrating FL and blockchain technology has the additional benefit of providing security and trust for the system, because all training events are recorded in a distributed ledger. Blockchain technology can address the threat posed by adversarial agents that try to tamper the training, a threat that is rarely addressed by the proposed FL methods in the literature \cite{Park2019}. Furthermore, blockchain can provide all the information needed to initialize the training process in the first genesis block \cite{Shayan2021biscotti}. This information includes the model structure, hyperparameters, and optimization algorithm.

Kim et al. propose \emph{BlockFL} \cite{Kim2020}, a blockchain-based decentralized FL architecture where agents send their local updates to so called miners (edge nodes with more resources) that share and cross-verify all the local updates before starting a Proof-of-Work (PoW). The miner who first finishes PoW receives a mining reward and creates a new block that is added to the blockchain. Each agent updates the local model from the most recent block, i.e., the global update is computed locally. While blockchain offers security, traceability and reliability, the substantial workload of the miners, comprising model verification, sharing and PoW, requires substantial computational, communicational and energy resources, as well as increases the latency of the training \cite{Nguyen2021}. In addition, there seems to be a lack of efficient cross-verification methods for the local model updates \cite{wang2021bcflsurvey}.

Shayan et al. propose Biscotti \cite{Shayan2021biscotti}, which is a privacy-preserving, blockchain-based fully decentralized FL system. Biscotti utilizes many cryptographic primitives to defend against model poisoning attacks and information leakage attacks: Multi-Krum defense prevents peers from poisoning the model, differential privacy protects shared model updates from inference attacks, and Shamir secrets provide secure aggregation. They propose their own Proof-of-Federation (PoF) consensus algorithm for generating the new block, which represents a single iteration of SGD. PoF consists of three stages: adding noise to the SGD updates, validating the updates, and securely aggregating the updates. Verifiable Random Functions (VRFs) are used to select a subset of peers responsible for the different stages of the PoF. The peers are selected proportional to their stake, which is the reputation that a peer acquires by positively contributing to the model training. 

Experiments with Biscotti show resilience against label-flipping poisoning attacks when 30\% of the peers are malicious \cite{Shayan2021biscotti}. They also show that Biscotti is able to converge under node churn (nodes joining and leaving/failing) that happens at the rate of 1 node joining and 1 node failing every 1.875 seconds. However, Biscotti has its deficiencies. It introduces a privacy-accuracy trade-off when differential noise is added to the SGD updates, which reduces the accuracy of the final model. Multi-Krum based model update validation scheme also has issues, because it needs to observe a large number of honest updates in each round, and it can also reject updates from peers that have non-IID data. In addition, Biscotti does not scale to large ANN model with millions of parameters due to communication overhead, and the stake based selection of the peers for the PoF stages does not account for peers that turn malicious after gaining enough stake in the system.

All the previously handled peer-to-peer learning methods assume full cooperation over the whole training time. In addition, the agents do not decide about whether they cooperate or not; the algorithms require them to exchange information at every round. With regard to our vision, we are interested in agents that are more self-interested and consider whether it is beneficial to cooperate. To this end, Yu et al. \cite{Yu2013repdiff} apply ATC diffusion strategy to learn a common target parameter among a network of self-interested agents. They formulate all the interactions between the agents as successive one-shot games. Each agent has a cost function that combines estimation accuracy and communication cost. At the beginning of each round, agents are randomly matched into pairs. Each agent calculates the adaptation step of ATC to obtain an intermediate result. Then, based on some prior knowledge that can be exchanged when the agents are paired, each agent evaluates the expected cost of sharing the intermediate result with the other agent. However, if the agents choose their action purely based on the minimization of instantaneous combined cost, the dominant strategy is not to share estimates. This is because the agents do not have a way to predict each other's actions, which results in a lack of belief in the other agent's actions.

To address the lack of belief, Yu et al. introduce a reputation scheme to allow an agent to assess the belief it has about the other agent's actions. Each agent holds a scalar reputation parameter that summarizes the history of the other agent’s actions as viewed by the agent itself. These reputation parameters can be used to predict whether the other agent will exchange information, and they also serve as incentives to share because if an agent chooses not to share when it is beneficial, the agent's reputation score is reduced. The sharing is beneficial when the improvement in the parameter estimation accuracy outweighs the cost of the communication. Using this reputation scheme, Yu et al. formulate a simple action-choosing policy for the agents. Their experiments with 10 agents show that the agents will cooperate in the beginning of the training, but will stop when the communication cost outweighs the improvement in the estimation accuracy. They also show that the convergence rate improves under their reputation scheme when compared to self-interested agents without a reputation scheme and to the fully cooperative agents.

Another thing to note is that the previous decentralized approaches target the convergence into a global consensus model. Vanhaesebrouck et al. take a peer-to-peer learning approach where the primary goal is for each agent to learn a personalized, local model \cite{vanhaesebrouck2017}. They present a collaborative learning setting, where agents have their own learning objectives and train their own local models while simultaneously interacting with a neighbourhood consisting of other agents with similar learning objectives. The core idea of collaborative learning is to smooth the local model parameters within the neighbourhood, while simultaneously preventing significant decreases in the local model accuracy as a result of the smoothing. A weight is assigned to each pair of neighbours to indicate the similarity between their local objectives.

Vanhaesebrouck et al. \cite{vanhaesebrouck2017} reformulate the collaborative learning problem into an equal \emph{partial consensus} problem and propose an asynchronous gossip algorithm based on ADMM for solving it. The partial consensus is formulated as a constraint requiring that two neighbouring agents agree on each other's personalized model parameters. Introducing such a constraint requires that each agent keeps a local copy of its every neighbour's model parameters. 

The ADMM based algorithm is computationally expensive and requires the introduction of many auxiliary variables. Bellet et al. \cite{bellet2018} propose a simpler and faster decentralized block coordinate descent algorithm for solving the same problem. They also address data privacy issues by applying differential privacy in the model updates sent between neighbours.

\subsubsection{Efficient Model Inference}

An important aspect to consider in distributed learning is how to ensure efficient and fast inference with the models. Resource-constrained nodes often face significant challenges when they need to calculate ANN model outputs. A large ANN model may not fit into a node's memory, not to mention the computational costs involved with executing ANN models. While cloud-based inference would remove these challenges, it does not serve the latency requirements of real-time applications, and raises issues of data privacy as the input data must be sent to cloud. Thus, methods for adapting ANN models to fit into the resource-constrained nodes are called for. This usually introduces trade-offs between inference accuracy and latency \cite{Xu2020}.

Adapting ANN models can start from designing new model architectures that require less from the hardware, or from compressing existing models through, e.g., low-rank approximations, knowledge distillation, pruning or parameter quantisation \cite{Xu2020}. Another viable option to reducing resource consumption and latency is conditional computation. This means that unnecessary calculations are turned off, as is done in early exit, where ANN model execution is stopped when a target output confidence level is reached \cite{Deng2019}. Model splitting, where a neural network is cut into parts that are executed by different nodes, is also a popular approach for model inference \cite{Deng2019}. However, finding a cut-point that best balances the issues of latency, privacy and resource consumption is not a trivial task \cite{Wang2020}. 


\subsection{Decision Making and Negotiation}\label{sec:decmaking}

\begin{figure*}
    \centering
    \includegraphics[width=0.85\linewidth]{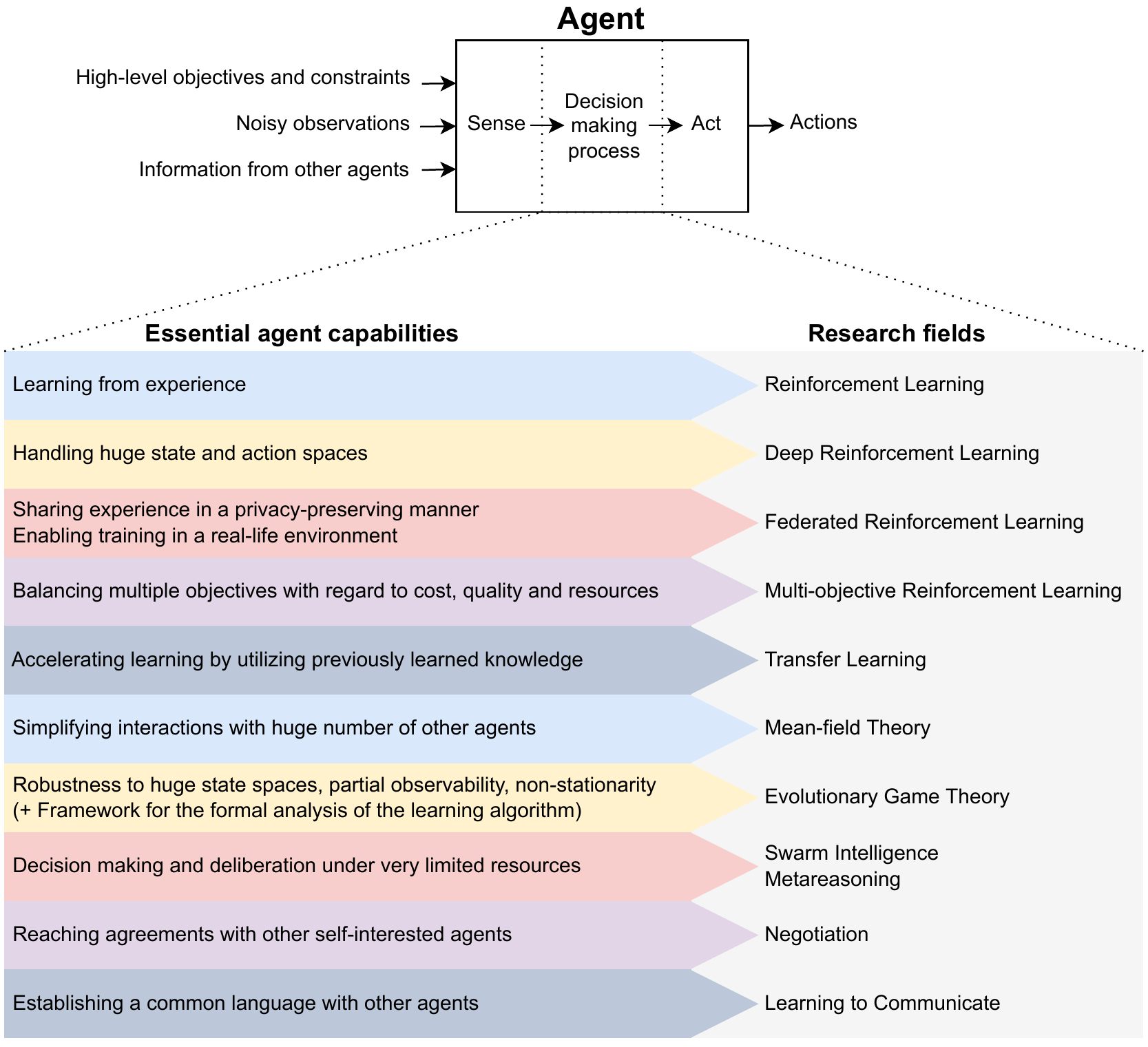}
    \caption{\textbf{Essential agent capabilities.} A computing continuum agent must have a set of essential capabilities in its decision making process. The figure highlights the most relevant capabilities an agent must have with regard to our vision, and points towards research topics that are crucial in fulfilling the capabilities.}
    \label{fig:dec_making}
\end{figure*}

In the highly dynamic computing continuum, decision making strategies\footnote{Words `policy' and `strategy' are used interchangeably in this survey} learned by the agents must be stable, but adaptive enough to conform to changes. Agents have to learn strategies based on their own local experience and interactions with neighbouring agents in order to make optimal decisions. 

In this section, we will focus particularly on sequential decision making methods that could lead to efficient solutions in the computing continuum orchestration. \emph{Multi-Agent Learning} (MAL) is a field that studies learning paradigms for MASs and develops algorithms to create adaptive, optimal decision making policies for agents. Reinforcement learning is the most commonly studied learning paradigm for a MAS \cite{Tuyls2012basics}. We believe that reinforcement based learning is a key ingredient for intelligent computing continuum orchestration, and hence, this section will mainly focus on reinforcement learning based sequential decision making.

Making a decision can be equivalent to reaching an agreement between self-interested agents, which requires negotiation techniques. For example, an agent can make the decision about offloading a task by itself based on, e.g., the current load on the node, task size and predictions about the changes in the load. However, the decision of where to offload the task is the outcome of negotiation among the autonomous, self-interested agents. An agent cannot make such a decision by itself, as it would violate the autonomy of the other agent. Hence, this section will also touch upon the current state of multi-agent negotiation.

Any communication between self-interested agents requires a common language. How this common language is established is a highly important research question, especially in the open computing continuum environment where different agents are designed and implemented by different people. Hence, we will also take a brief look at AI methods that enable learning a shared language.

Overall, the type of agents we envision to be acting in the computing continuum must have certain essential capabilities in their decision making processes. These capabilities are depicted in \cref{fig:dec_making} along with the research fields that are central for developing solutions that can fulfill these capabilities. The following subsections will introduce the state of the art in each of the fields.

\subsubsection{Reinforcement Learning}
\textbf{Fundamentals.}
\emph{Multi-Agent Reinforcement Learning} (MARL) environment is usually formulated as a stochastic game, also known as a Markov game \cite[p. 160]{shoham2008book}. A Markov game is formally defined as a tuple $(\mathcal{N}, \mathcal{S}, \mathcal{A}, T, R )$, where $\mathcal{N}$ is a finite set of $K$ agents, $\mathcal{S}$ is a finite set of states, $\mathcal{A} = A_1 \times \dots \times A_K$ is the joint action space with $A_k$ being the action space of agent $k$, $T: \mathcal{S} \times \mathcal{A} \times \mathcal{S} \rightarrow [0, 1]$ is the transition probability function, 
and $R = (R_1, \dots, R_K$), where $R_k: \mathcal{S} \times \mathcal{A} \times \mathcal{S} \rightarrow \mathbb{R}$ is a real valued reward function for agent $k$.

MARL generally recognizes three different settings for the interaction between the agents: fully cooperative, fully competitive and mixed setting \cite{zhang2021multi}. In fully cooperative settings, agents have a shared goal and hence, each agent usually also has the same reward function: $R_1 = \dots = R_K$. The goal in fully cooperative settings is to maximize the common return. Fully competitive settings are formulated as zero-sum games, where two agents have opposing goals, i.e., $R_1 = -R_2$. Mixed settings are formulated as general-sum games, where there are no restrictions on the goals and relationships among the agents; agents are \emph{self-interested} and their individual rewards can be in harmony or in conflict with each other.

Agents interact with the environment in discrete time steps. Each agent follows a policy that maps each state-action pair to the probability of selecting the action in the state. The state-value function of agent can be defined as the expected sum of discounted rewards that is obtainable from the state \cite{zhang2021multi}. The goal of each agent is to find an optimal policy that maximizes the state-value function. However, the expected return of an agent is influenced by the actions of the other agents, as the reward an agent receives is dependent on the joint action of the agents. In the case where the other agents are also learning agents with non-stationary policies, the environment itself becomes non-stationary. This opponent-induced non-stationarity is a particularly severe challenge in the computing continuum, as it violates the fundamental assumption of the Markovian property (a stationary environment) behind single-agent RL algorithms \cite{hernandezleal2019survey}. Thus, algorithms developed for finding optimal policies in the single-agent settings should not be applied as is in multi-agent systems, because all derived guarantees are lost \cite{Tuyls2012basics}.

MARL algorithms developed for finding the optimal policy for each agent need solution concepts to address the dependence of an agent's optimal strategy to the strategies of the other agents. \emph{Nash Equilibrium} (NE) is a common choice for the solution concept. An NE for a Markov game is defined as a joint policy where no agent can improve their expected return by unilaterally deviating from the joint policy \cite{zhang2021multi}. In other words, the policy of agent is its best response to the joint policy of the other agents. Most of the MARL algorithms aim to converge to a NE, as it is a stable point from which none of the agents has any incentive to deviate. 

Despite its prevalence, NE has deficiencies as a solution concept. It is not often unique, and it does not guarantee optimality, because it assures that no single agent can unilaterally improve their reward, not that the global reward is maximized or that there is no other equilibrium where agents could simultaneously improve their rewards \cite{nowe2012chapter}. In addition, convergence to NE is a reasonable goal only under the assumption that agents are perfectly rational and capable of infinite mutual modelling \cite{zhang2021multi}. However, this is not a reasonable assumption in many real-life situations, such as when agents with bounded rationality (e.g., computationally limited, which is a relevant characteristic of agents in the computing continuum) are involved, because such agents are only able to do finite mutual modelling. These situations call for different solution concepts to pursue as the main goal of MARL algorithms.

One very important aspect besides the stability of a MARL algorithm is the capability to adapt to the strategies of other agents. In its extreme form, this is manifested in the so called agent tracking algorithms for mixed settings, in which agents follow and observe the actions of others to build models about their strategies, and then aim to provide their best response to these models rather than focusing on having a stable strategy \cite{bucsoniu2010multi}. However, a practical algorithm needs to find a balance between the \emph{stability} and \emph{adaptivity} of a decision making strategy. In other words, algorithm should detect when the change in the other agents' behavior is more permanent, and not react to every little change in behavior that may be a result of random exploration.

Developing MARL algorithms is very challenging, because in addition to opponent-induced non-stationarity, there is a plethora of other challenges to consider. These issues include balancing exploration-exploitation trade-off, sample inefficiency (it takes an incredible amount of samples and time to train an adept agent), sparse rewards (every transition does not return a feedback), dependability (in the sense of, e.g., guaranteeing a certain level of average reward), coordination, communication efficiency, preventing the overfitting of agents, handling dysfunctional or malicious agents, credit assignment problem (in fully cooperative settings), state and action space explosion (scalability) and partial observability. In this survey, we are mainly focusing on the core issues of opponent-induced non-stationarity, scalability and partial observability.

From theoretical perspective, the introduced Markov game framework forms the theoretical foundation of MARL along with the extensive form game framework \cite{zhang2021multi}. In Markov games, agents choose their actions simultaneously and receive an immediate reward, whereas in extensive form games, agents choose their actions sequentially and receive a reward only at the end of an action sequence. Markov game framework can handle only the fully observable environment, whereas extensive form framework is able to handle the partially observable environment. Extensive form game framework is generally used in non-cooperative (fully competitive or mixed) settings \cite{zhang2021multi}. However, extensive form games are not always a good model for large or realistic multi-agent settings \cite[p.147]{shoham2008book}.

How to handle partial observability, that is, not being able to observe the exact environment state, or the actions and/or rewards of the other agents, is an important question, because it is not often realistic to assume full observability in practical MARL applications already due to the noise in sensor observations and communication channels. It is possible to extend Markov game framework to the partially observable case. The resulting model is called Partially Observable Stochastic Game (POSG) \cite{Hansen2004posg}, which is formally defined as a tuple $(\mathcal{N}, \mathcal{S}, \mathcal{A}, \mathcal{O}, \mathcal{P}, R, b_0)$, where $\mathcal{N}$, $\mathcal{S}$, $\mathcal{A}$, and $R$ are defined as in Markov games, $\mathcal{O}=O_1 \times \dots \times O_K$ is the set of joint observations with $O_k$ being the finite observation set of agent $k$, and $\mathcal{P}$ is a set of state transition and observation probabilities with $\mathcal{P}(\hat{s}, o | s, a)$ being the probability that taking the joint action $a$ in state $s$ leads to transition to state $\hat{s}$ and joint observation $o$. $b_0 \in \Delta(\mathcal{S})$ is the initial state distribution at time $t = 0$. Policy now maps action-observation histories to actions.

In fully cooperative settings, POSG reduces to decentralized partially observable Markov decision process (dec-POMDP) \cite{Oliehoek2016}, where all agents have the same reward function. However, even in the fully cooperative case, partially observable models are very difficult to solve, because it has been proved that solving decentralized POMDPs is NEXP-complete \cite{Bernstein2002decpomdp}. Most MARL algorithms proposed for partially observable problems consider only fully cooperative settings, and they require some form of centralization so that the decentralized problem can be reformulated as a centralized POMDP \cite{zhang2021multi}. Reformulation as a centralized model can also be achieved by communication, for example, if each agent broadcasts their observation, they all receive the joint observation and can compute joint beliefs, which basically turns the decentralized problem into a centralized POMDP.

Note that the traditional definition of POSG is underspecified, because it does not include the specification of the communication capabilities of the agents. Oliehoek and Amato introduce Multiagent Decision Problem (MADP) to address this issue \cite[p. 25]{Oliehoek2016}: they specify two models, environment model and agent model. The previously introduced POSG model can be interpreted as a Markov multiagent environment model, which is underspecified in that it does not specify the information on which the agents can base their actions, or how they update their information. This can be made explicit with an agent model, which for each agent $k$ is a tuple $(\mathcal{I}_k, I_k, A_k, O_k, \mathcal{Z}_k, \pi^k, \iota_k)$. Here, $\mathcal{I}_k$ is a set of information states, $I_k$ is the current information state of the agent, $A_k$ and $O_k$ are the action and observation sets of the agent, $\mathcal{Z}_k$ is the set of auxiliary observations that can be obtained through, e.g., communication, $\pi^k$ is the (stochastic) policy of the agent, and $\iota_k: \mathcal{I}_k \times A_k \times O_k \times \mathcal{Z}_k \rightarrow \Delta(\mathcal{I}_k)$ is information state function or belief update function.

In MADP, the problem designer must specify optimality criterion (such as the expected sum of discounted rewards), the environment model, and a subset of the elements of the agent model. The algorithm developed for solving the problem must optimize the non-specified elements of the agent model, in order to maximize the value given by the optimality criterion. We believe that the inclusion of this type of agent model is particularly relevant in the future computing continuum orchestration solutions, as it characterizes the knowledge and capabilities of an agent, allowing a more local, agent-centric view on the decision problem.

\textbf{Algorithms.}
Hernandez-Leal et al. provide a useful categorization of the proposed MARL algorithms based on how they address the opponent-induced non-stationarity \cite{hernandezleal2019survey}. They identify five categories with increasing sophistication in the approach to handling the non-stationarity: ignore it, forget it, respond to target opponent models, learn the opponent models, or rely on a theory of mind. Ignoring the opponent-induced non-stationarity has been a popular approach due to its simplicity and easiness: agents are treated as \emph{independent learners} in an environment where opponent-induced non-stationarity is regarded as stochastic noise within the transition model, which enables the use of algorithms developed for single-agent learning, such as Q-learning. In other words, algorithms that ignore non-stationarity learn a stable policy while assuming that other agents are using stationary policies, i.e., are not learning. Obviously, if opponents\footnote{Similarly to Hernandez-Leal et al. \cite{hernandezleal2019survey}, we are using the word 'opponent' to refer to another agent in the environment regardless of whether the objectives of the agents are aligned or not.} are learning or change their strategy, these algorithms will fail. We do not believe that this approach will be a sustainable and feasible approach in the computing continuum orchestration.

Algorithms using forget approach are typically model-free algorithms that adapt to the changing environment by updating the strategies with the most recent information while forgetting old information. A representative example of algorithms in this category is WoLF-PHC \cite{bowling2002}. WoLF-PHC is an algorithm for mixed settings that does not require observability of the actions or rewards of other agents and is based on using Win-or-Learn-Fast (WoLF) principle in policy updates and Q-learning in Q-function updates. It handles non-stationarity by adjusting between two learning rates based on whether agent is interpreted to be winning or losing. The interpretation is determined by comparing the current policy’s expected payoff with that of the average policy over time. WoLF-PHC can be used in cases where agents have heterogeneous learning strategies, but it assumes that the opponents are slowly adapting. This type of adaptation based on the newest information and the notion of the current status (such as winning or losing in WoLF-PHC) while forgetting old information could possibly lead to some orchestration solutions among the more resource-constrained agents that cannot be expected to uphold complex models about the other agents or extensive histories of their interactions with the environment.

Algorithms that respond to target models optimize against clear and defined opponent strategies. There exist many algorithms in this category due to the fact that it is easier to give guarantees against specific opponents than general classes \cite{hernandezleal2019survey}. Replicator Dynamics with a Variable Learning Rate (ReDVaLeR) \cite{Banerjee2004redvaler} is a good example of algorithms in this category. It is a model-free algorithm for mixed settings, which provably learns a stationary best response against stationary opponents and NE against adaptive opponents that \emph{all} use the same ReDVaLeR algorithm (self-play). In addition, ReDVaLeR provides a constant bound on expected regret, which makes the algorithm robust against opponents that use arbitrary non-stationary policies. However, it assumes that an agent can distinguish between self-play and otherwise non-stationary agents, requires that opponent actions are observable, and has been developed for repeated games (single state).

Many algorithms in the respond to target opponent models category have been either directly developed for or evaluated only in 2-player competitive games, and they do not often address in any way what happens when an agent encounters other agents that do not belong to the target classes. However, these algorithms can offer noteworthy solutions that could be useful in computing continuum orchestration solutions, such as the guarantee of a bounded regret against opponents using arbitrary non-stationary strategies in ReDVaLeR algorithm. In the open computing continuum environment it is unrealistic to assume that all agents are using the same learning algorithm; guaranteeing a certain level of performance against other agents using any type of learning algorithms is desirable. 

Algorithms that learn opponent models form strategies based on how the opponent is behaving and aim to adapt the models to changes in the opponent behavior. That is, the algorithms update the opponent models and the current strategy constantly to keep up with the non-stationary opponents. A representative example of algorithms in this category is BPR+ \cite{Hernandez2016bpr}, which has been developed for a repeated game where the opponent switches among several stationary strategies in a way that it can either start using a completely new strategy or return to a previously used one. BPR+ learns all models through interactions, and stores previous models to memory when it detects a change in opponent's behavior, after which it starts learning a new model. The opponent behavior is modelled with an MDP. Thus, it needs to be able to observe opponent actions, can require a lot of memory resources and is not a very scalable algorithm.

Theory of mind algorithms are the most evolved ones in regard to handling non-stationarity as the models assume that the opponent is also establishing opponent models, whereas algorithms in learn models category do not take into account whether the opponent is modelling them or not. These kind of algorithms naturally require high computational resources to execute as they aim to perform complex, recursive and strategic reasoning. Thus, majority of the work studying this approach have been done in the context of one-shot, stateless games \cite{hernandezleal2019survey}. Interactive POMDP (I-POMDP) \cite{Gmytrasiewicz2005ipomdp} is perhaps the most known framework in this category for sequential decision making. It extends POMDP model to distributed multi-agent systems by incorporating models of other agents into the state space.

Learn opponent models and theory of mind approaches could only be deployed by agents with basically unlimited resources due to their high computational and memory load. Furthermore, the approaches do not easily scale up to a large number of agents. Theory of mind approaches have been mainly used for predicting the outcome of one-shot games, and they cannot usually be used for online learning. For example, I-POMDP requires that the transition and observation probabilities are known. Some work that aims to enable online learning with I-POMDPs \cite{Ng2021} and increase the scalability of I-POMDPs \cite{sonu2015ipomdp} does exist.

\subsubsection{Deep Reinforcement Learning}
Computing continuum environment modelled as a MAS consists of a large number of agents, which induces a scalability issue for MARL algorithms. This is due to the combinatorial nature of MARL: joint action space dimension increases exponentially with the number of agents. Many MARL algorithms are based on value or policy iteration, which involves tracking values for each state or state-action pair. This is obviously not possible when the dimensionality of state and joint action spaces grows. One solution to the scalability issue is function approximation: rather than tracking the state value or state-action value functions in tabular form, these functions are approximated with, for example, ANNs. Combining deep learning with MARL has resulted in the formation of the field \emph{Multi-Agent Deep Reinforcement Learning} (MADRL), the algorithms of which have been particularly successful in vision-based domains, such as games \cite{hernandezleal2019critique}. 
However, the main disadvantage of these methods is that providing theoretical guarantees about convergence of algorithms in the case of function approximations is extremely difficult for non-linear ANN approximations \cite{zhang2021multi}.

Using ANN models to approximate value functions brings the benefit of better generalization across states and removes the need to manually handcraft features that are used to represent state information. However, the problems of non-IID training data and divergence from optimal values pester the use of ANN approximators \cite{hernandezleal2019critique}. Many single-agent DRL methods that are based on the seminal method Deep Q-Network (DQN) \cite{Mnih2015DQN}, use experience replay memory to break the ties between the highly correlated sequential training samples (state-action pairs), and address the divergence issue by adopting two NNs to approximate value function so that the weights of the other, called target network, are updated only periodically. Transferring experience replay into a MAS is not straightforward because past observations become obsolete due to the adaptation in agents' policies. The two main approaches that have been proposed to overcome this are fingerprinting to disambiguate the age of the samples, and leniency values, which are basically temperature values that gradually decay by the number of state-action pair visits \cite{hernandezleal2019critique}.

Multiple MADRL algorithms have been proposed in the literature to overcome the same type of challenges that traditional MARL algorithms encounter \cite{hernandezleal2019critique, Nguyen2020survey, Gronauer2021survey}. One interesting method that aims to overcome the opponent-induced non-stationarity is Deep Loosely Coupled Q-network (DLCQN) \cite{castaneda2016DLCQN}. DLCQN is based on the loosely coupled Q-learning \cite{Yu2015MultiagentLO}, where each agent has an independence degree for each state, giving the probability that the agent can act independently in a given state. This degree is adjusted whenever a negative reward is received. However, loosely coupled Q-learning was originally developed for 2-agent robot coordination problems, and DLCQN does not extend this to a bigger number of agents. Nevertheless, the notion of an independence degree simplifies the decision making of an agent in the states where it can assume to be relatively independent of other agents. Studying the applicability of such a notion in computing continuum orchestration problems could be beneficial.

In MADRL algorithms, partially observable environments are usually handled by utilizing deep recurrent neural networks, such as Long Short Term Memory (LSTM) networks. Solutions that aim to overcome both, non-stationarity and partial observability, utilize some form of communication or centralization to share information and coordinate the actions of the agents. Currently, \emph{Centralized Training with Decentralized Execution} (CTDE) is the standard, most widely adopted paradigm when it comes to learning in a partially observable, non-stationary MAS \cite{zhang2021multi, Gronauer2021survey}. It involves sharing information (e.g., observations and actions) during training through a centralized entity. This entity can provide the information for all the agents so that they can optimize their policies, or the central entity may also calculate the optimal policy. During execution, all this extra information is discarded, and an agent simply follows the policy it calculated during training or received from the central entity after the training. CTDE is particularly popular in algorithms developed for solving fully cooperative dec-POMDPs, where it simplifies the theoretical analysis \cite{zhang2021multi}. CTDE works mainly in applications where the learning can be done a simulator or a laboratory, where there are no communication constraints and extra state information is available \cite{Nguyen2020survey}.

CTDE in its current state is not a very suitable paradigm for learning policies in the mixed settings of the computing continuum environment. Agents cannot use the policies learned in a limited, simulated environment as is in the dynamic computing continuum. Furthermore, agents should be able to adapt their policies if during execution they note that their performance with the current policy deteriorates. That is, they should be able to go back into the training phase, which in CTDE algorithms would require access to all the extra information that was available during the initial training in a simulator. Hence, training paradigm in the computing continuum must emphasize decentralization with the ability to communicate with the local neighbourhood, while taking into account that this communication can be intermittent and limited. 
However, we do recognize that policies learned in a simulated environment could provide prior information for learning in the real environment (i.e., transfer learning aspect, which will be discussed later).

Mixed setting is a notoriously hard setting when it comes to developing provably convergent algorithms even in the MARL domain \cite{zhang2021multi}, not to mention when utilizing non-linear ANN approximations. Mixed setting in RL is generally underexplored when compared to fully cooperative or fully competitive settings. We see mixed setting MADRL algorithms as an important research direction for developing solutions in computing continuum orchestration. 

\subsubsection{Federated Reinforcement Learning}
Training accurate ANN models requires a substantial amount of training data, and it would take from one agent a long time to collect enough observations in the environment to attain enough data for training accurate ANN models. Furthermore, the observation range of one agent is limited to its local view. Sharing observations between agents is one way to increase the sample efficiency and help one agent to attain a more global view on the environment state, but it is not always possible due to data privacy and communication efficiency issues. 

Combining FL with MARL opens up the possibility for a group of agents to collaboratively learn ANN models (actor and/or critic models) by periodically exchanging encrypted ANN model parameters without actually sharing any observations \cite{qi2021federatedRL}. The studies on the combination of FL and RL are currently at scarce particularly in the type of real-life settings encountered in the computing continuum, that is, a setting where agents share the same environment while having their own sets of observations and actions. We see federated RL as a promising research direction for enabling MARL training in real-life settings (basically bringing CTDE into a real-life environment). Furthermore, the possible applicability of decentralized FL methods (See \cref{sec:declearning}) in RL settings is another open research direction, which could take us towards a decentralized training paradigm.

Even though FL allows sharing experience in a privacy preserving way, it has the disadvantage of confining each agent to using the same, homogeneous ANN structure. To address this issue, Federated Reinforcement Distillation has been proposed \cite{cha2019FDR}, where each agent shares their local proxy experience memory with an aggregator that creates a global proxy experience memory. Each agent then aims to minimize the cross entropy loss between their local policy and the policy of global proxy experience memory. An improved version that uses mixup data augmentation algorithm to interpolate the global experience replay memory has been proposed as well \cite{Cha2020augFDR}.

\subsubsection{Multi-objective Reinforcement Learning}
RL methods usually aim to optimize a single objective. However, any orchestration solution in the computing continuum must inherently consider multiple objectives with regard to cost, quality and resources. Multi-Objective Reinforcement Learning (MORL) is a field that aims to develop methods for RL problems where multiple different objectives must be optimized. This means finding an appropriate trade-off between competing objectives. A central solution concept in MORL for defining the set of best trade-off solutions is \emph{Pareto efficiency} \cite{VanMoffaert2014paretoQ}: the solution for the multi-objective problem cannot be Pareto dominated by any other solution, meaning that there cannot be any other solution where some objectives will be improved and no objective will be deteriorated. The set of such non-dominated, mutually incomparable solutions is called \emph{Pareto front}.

In MORL, instead of a single scalar reward, agents receive a reward vector where each component measures the performance on a different objective. Majority of MORL approaches handle the multiple objectives with scalarisation. This means that the reward vector is mapped into a scalar value with a scalarisation function, which most typically is a linear weighted sum of the reward vector components \cite{Radulescu2020MORLsurvey}. After this, methods developed for single-objective RL problems can be used. However, scalarisation has multiple disadvantages. It requires a priori information about the preferences over different objectives, such as whether low latency is more preferred than low cost. Further, if the preferences change significantly over time, the policy must be re-learned with the new reward function. It is also important to note that non-linear scalarisation may in some cases break the applicability of single-objective RL methods \cite{Roijers2013MORL}, and that linear scalarisation is known to find solutions only from the convex regions of the Pareto front \cite{Vamplew2008}.

In many situations, a more desirable way to handle the multiple objectives is to find a set of policies that properly approximates the Pareto front, which is referred to as a multi-policy approach \cite{VanMoffaert2014paretoQ}. In this way, it is easier to evaluate the different trade-offs that can be done between the different objectives, and create a selection mechanism that can choose the policy for execution based on the current preferences over objectives. In principle, this could be done with the scalarisation approach by running the learning algorithm with different scalarisations. However, choosing the scalarisations so that the resulting policies are properly spread in the objective space is difficult \cite{vanMoffaert2014weight}, and running the same algorithm several times with different reward functions in the computing continuum is infeasible. Hence, the MORL algorithms for computing continuum should learn a set of optimal policies in a single run. 

One example of an algorithm that can provide a set of optimal policies in a single run is Pareto Q-learning \cite{VanMoffaert2014paretoQ}, which, however, has been developed for a single agent setting. In general, MORL applied in a multi-agent setting has not been comprehensively explored. This line of research is essential for the future computing continuum orchestration solutions, as balancing trade-offs between different objectives is unavoidable. How to efficiently implement learning in a MAS with multiple objectives, while taking into account adaptation to other agents, changing preferences, and limited resources is an open research question.

\subsubsection{Transfer Learning}
Transfer learning aims to accelerate the learning process by reusing knowledge. Transfer learning methods make it possible to transfer knowledge from powerful, well-trained agents to the agents that are just starting the learning procedure or that do not have enough time, memory or computational resources to train as accurate models. For example, agents operating in the cloud have basically unlimited resources for reasoning and learning, and their knowledge could be utilized by the resource-limited agents operating on the edge. 

Furthermore, learning a policy from scratch in a real environment is not feasible due to the huge amount of experience needed to train a policy, and because in the early stages of learning, agents can make very awful decisions. Hence, knowledge obtained through, for example, training in a simulated environment or from demonstrations by human experts could offer an excellent starting point for further learning in the real environment. 

Generally, transferring knowledge has been studied from three different perspectives: an agent can reuse its own knowledge from previous tasks, agent can gain knowledge from observing other agents, or agent can get direct advice from experts \cite{Silva2019survey}. The first aspect is referred to as intra-agent knowledge transfer, whereas the latter two belong to the inter-agent knowledge transfer. 

In terms of the intra-agent knowledge transfer, one interesting paradigm is curriculum learning, where the idea is to divide a problem into subproblems of varying difficulty level. Then starting from the easier tasks, agent transfers knowledge across progressively harder tasks. These approaches are often more concerned with how to define the curriculum, i.e., a task sequence, rather than how to reuse knowledge, since all tasks inside a curriculum are assumed to be inside the same domain \cite{Silva2019survey}. In this regard, Silva and Costa propose a method for automatically generating a curriculum based on object-oriented task descriptions \cite{Silva2018curriculum}.

The majority of the transfer learning literature is focused on the inter-agent knowledge transfer \cite{Silva2019survey}. These approaches are based on the assumption that all tasks are in the same domain, and knowledge is transferred through either explicit or implicit communication. These methods can be direct action suggestions from other agents (usually in fully cooperative settings), transferring human expertise, receiving heuristics or reward shaping from teachers to improve exploration, learning from demonstration or imitating other agents through observing them. There is not yet a lot of work that consider knowledge transfer for agents learning with deep networks; however, there are works indicating that unprincipled knowledge transfer in MADRL can lead to more catastrophic negative transfer effects than in the traditional MARL \cite{Glatt2016, Silva2019survey}. In addition, the security aspect of knowledge transfer between agents has not been considered practically at all, that is, agents participating in the knowledge transfer are assumed to be benevolent.

Transfer learning is among the most important research topics for the future computing continuum orchestration, because the ability to reuse knowledge is a corner stone of the type of intelligent, adaptive and autonomous agents considered in our vision. Current research is heavily dependent on human efforts and validates proposed approaches mainly in simulations of simple problems \cite{Silva2019survey}. However, developing solutions for open and complex real-life environments such as the computing continuum is extremely hard. Difficulties already start from such a fundamental issue as knowledge representation, that is, the transfer of knowledge between the agents relies on the fact that the agents have the same way of representing knowledge. To which extent such an assumption is valid among the heterogeneous agents of the computing continuum is an open research question.

\subsubsection{Mean-field Theory}
As previously stated, computing continuum entails a large number of agents. Most of the approaches for reinforcement learning in a MAS are designed for or validated in settings with a maximum of tens of agents. Many methods that aim to model other agents require a separate model for each of the agents in the environment. This very quickly becomes infeasible when the number of agents grows, not to mention when also considering agents with limited resources.

Solutions for this scalability issue could be searched with \emph{mean-field game theory}. Mean-Field Games (MFGs) are designed for situations with possibly thousands of agents. In MFG, an N-player game is reformulated as a 2-player game, where each agent is playing against a virtual opponent, which presents an average player of the entire (homogeneous) population. In optimal control theory, where an agent population is modelled as a controlled stochastic dynamical system, solving MFGs involves solving two coupled partial differential equations, which are usually approximated with neural networks \cite{Caines2017mfg}.

Studying mean-field theory in MARL has only recently gained attention \cite{zhang2021multi}. One notable work is by Yang et al. \cite{yang2020mfmarl}, who introduce a mean-field Q-learning and mean-field actor-critic algorithms for non-cooperative settings. In their approach, an agent aims to find the best response against the mean action of its neighbourhood (which approximates the joint action in Q-function update), i.e., it needs to be able to observe the actions and rewards of its neighbours. Further research is needed to study the validity of mean-field theory in MARL, especially in the type of real-life settings that include partial observability.

\subsubsection{Evolutionary Game Theory}
Combining evolutionary game theory concepts with RL opens many interesting possibilities. The benefit of evolutionary game theory over traditional game theory is the fact that it does not require agents to be hyper-rational players, and an equilibrium can change dynamically \cite{tuyls2005evolutionary}. Evolutionary game theory has even been proposed as the preferable framework for studying multi-agent learning formally \cite{Bloembergen2015survey}. This stems from the concept of \emph{replicator dynamics}, which can be utilised in the game theory to describe an agent's strategy change over time as a game is repeatedly played and a policy iteratively updated. For MARL algorithms, a link between replicator dynamics and RL can be used to derive mathematical model of the infinitesimal time limit of various learning algorithms, i.e., a model of learning dynamics (predicting the behaviour of an algorithm in the limit). These kind of dynamical models are very useful in hyperparameter tuning or in the development of new learning algorithms \cite{Bloembergen2015survey}. Also, ``the relative fitness" of different strategies can be compared within a population, dependent on the frequencies of those strategies in the population.

Evolutionary algorithms (such as genetic algorithms, evolutionary programming, genetic programming and evolutionary strategies) have been utilized in reinforcement algorithms that search the space of policies \cite{moriarty1999}. The main idea is to start with the set of policies and evolve them until a policy with good enough fitness is reached. Evolutionary Algorithm Reinforcement Learning (EARL) methods differ mainly on how the policies are represented, and whether the policy is seen as one `chromosome' that is evolved as a whole or whether it is divided into parts.

Evolutionary algorithms are an extremely promising direction for complex MARL applications, as they can address several fundamental challenges in RL problems, such as large state spaces (though generalization and selectivity in policy representations), partial observability (if the change in environment is slow) and non-stationarity. However, they are not very suitable for online learning due to the requirement of large number of experiences, and the fact that some generated policies may perform very awfully. Thus, they should be trained in simulation models of the environment. In addition, EARL algorithms do not preserve information about bad states or decisions, and rarely visited states can drift to random, possibly bad actions due to mutations \cite{moriarty1999}.

\subsubsection{Swarm Intelligence}
We have so far focused on the MARL paradigm that aims to make every agent a highly intelligent decision maker while assuming that the agents are fully rational. This, however, requires considerable resources from a single agent. In the hierarchy of the computing continuum, the lower the agents are, the less resources they have for running decision making algorithms. These agents should still be able to make coordinated decisions about, for example, optimally allocating their energy and communication resources so that they can transmit messages while minimizing the used transmit power and the interference they cause to each other. Such coordinated decision making should happen in a decentralized manner through local interactions. We regard Swarm Intelligence (SI) as a key MAL paradigm for developing self-organizing, decentralized and adaptive decision making solutions for agents with limited capabilities.

SI investigates the emerging behavior of a (large) group of simple, rationally bounded agents that achieve complex, intelligent behavior through their interactions \cite{Tuyls2012basics}. An inherent assumption in many SI methods is that the agents are fully cooperative, which can also be a reasonable assumption for the low-level agents in the computing continuum at least inside the same administrative domain. An agent with very limited resources does not have enough capacity to learn complex behaviors as a unit, as is the purpose in MARL approaches, but rather, the complex behavior must be an emergent property of a group of such agents interacting on a local level.

SI has been especially studied for combinatorial optimization, and a plethora of SI optimization algorithms have been proposed in the literature \cite{Chakraborty2017}. SI methods have also been investigated inside multiple different domains, such as robotics, Unmanned Aerial Vehicles (UAVs) and IoT \cite{Schranz2020, Zhou2020swarm, raslan2020}. Our purpose is not to provide a comprehensive analysis of the SI research, but rather briefly highlight a couple of concepts that we regard particularly interesting in terms of the computing continuum orchestration solutions, namely \emph{well-informed agents} and \emph{adaptive personality traits}.

When there is a group of simple agents sensing the conditions in an environment and trying to reach a coordinated decision, it is easy to imagine that some agents are more privileged than others; their observations can be less noisy and more accurate, or they may have access to information that the other agents do not have. These well-informed agents should have more impact on the behavior of the group. Such an idea is implemented, for example, in the simple, decentralized decision making algorithm proposed by Yu et al. \cite{yu2010implicit}, where the actions of the whole group of agents are guided by the actions of a few well-informed agents. The informed individuals are not explicitly assigned as leaders; the algorithm is based on implicit leadership, meaning that the well-informed agents have a positive confidence factor about their information state, which influences the behavior of the rest of the group.

Adaptive personality traits is a concept that has been utilized particularly in swarm robotics \cite{Ding2002, Givigi2006}. The idea is to have a set of personality values that affect the action choice. These traits of personality are adapted based on reinforcement learning, that is, traits that lead to a positive feedback are enforced while traits with a negative feedback are weakened. Utilizing personality traits for simple agents could lead to novel solutions in the computing continuum orchestration. For example, they could offer an efficient way to implement the self-interested nature also for the resource-constrained agents; a mixture of two personality traits, cooperative and non-cooperative, determines whether an agent is more inclined to take an action that benefits its own utility or an action that benefits the collective utility.

\subsubsection{Metareasoning}
Bounded rationality is an inherent characteristic for majority of the agents in the computing continuum due to the limited amount of resources they have for reasoning about decisions. A promising research direction for handling the bounded rationality of the resource-constrained agents in the computing continuum is metareasoning. The idea of metareasoning is to monitor the decision making process of an agent to enable it to make `good enough' decisions in situations where there is not enough time/resources for thorough deliberation \cite{Langlois2020metareasoning, russell1991metareasoning}.

There exists different metareasoning structures for a MAS, which differ in terms of whether metareasoning is part of an operating agent or an agent of its own controlling the operating agents, and whether the metareasoning processes of different agents coordinate with each other or not. Metareasoning approaches have been applied in multiple different types of problems, such as coordination, planning and scheduling, communication, resource allocation, belief updating and task delegation \cite{Langlois2020metareasoning}.

Metareasoning approaches also differ in the way how the meta-level processes control the object-level processes (i.e., how the monitoring algorithms control the decision making algorithms). They can stop the execution of an object-level algorithm (in the case where the decision making algorithms are so called anytime algorithms \cite{Svegliato2018anytime}), modify parameters or reasoning rules of object-level algorithms, select an object-level algorithm for execution, determine whether it is beneficial for an object-level process to communicate or coordinate with others (and how), provide information about other agent's reasoning strategies (that it has deduced), or determine whether it is beneficial for agents to redefine their relationships (in hierarchical settings, e.g., master-slave) \cite{Langlois2020metareasoning}.

\subsubsection{Negotiation}\label{sec:negotiation}
Reaching agreements between agents with conflicting interests is an essential part of the envisioned computing continuum orchestration. Depending on the problem at hand, this negotiation can be bilateral (one-to-one) or multilateral (one-to-many, many-to-many), and it can concern a single issue or multiple issues. Regardless of the number of participants or issues, negotiation in the computing continuum will always be an incomplete information game: agents do not initially know what the utilities and strategies of the other agents are.

Automated negotiation has been widely studied in the literature (see e.g. \cite{Kraus1997, Jennings2001neg, Cosmin2010, Fatima2014, Lopes2014, Baarslag2016, Kiruthika2020}). Generally, a negotiation mechanism has three main components \cite{Ren2004, Jennings2001neg}: 1) the set of possible outcomes; 2) negotiation protocol that defines the rules for the negotiation; 3) the negotiation strategies of the participants that, given the set of possible outcomes and the negotiation protocol, define how the agents decide about their actions during the negotiation. There exist a plethora of negotiation protocols in the literature that differ in terms of, e.g., the number of participants, the number of issues, the way multiple issues are handled (e.g., negotiated independently in parallel, as a package deal, or sequentially), time constraints, the type of setting (cooperative, competitive or something in between), information state (complete, incomplete), and whether limited communication or computation is considered.

We do not believe that there exists one universally correct or suitable approach on negotiation in the computing continuum. Computing continuum poses many challenges that affect the details of the deployed negotiation mechanism. For example, negotiation over administrative boundaries can be considered much more competitive than negotiation inside the same administrative domain, as the agents inside the same administrative domain ultimately aim to reach a globally optimal system state within that domain. Further, agents higher in the hierarchy of the computing continuum have more resources for their negotiation mechanism than the agents on the lower levels. However, any deployed negotiation mechanism in the computing continuum should address the problem of incomplete information, enforce the loose coupling and autonomy of the agents, and be secure due to the openness of the system and the self-interested nature of the agents. Security requires designing trust and reputation mechanisms so that the self-interested agents can better assess with whom to interact and are incentivized to be truthful in their interactions (see e.g. \cite[pp. 381-414]{Weiss2013}).

In the scope of this article, due to the overwhelming amount of negotiation related work, we are focusing on giving a brief overview of two research themes that we see particularly important for negotiation in the computing continuum regardless of the actual details of the negotiation situation: opponent modelling and argumentation. 

A simple way to handle the incomplete information of the participants in the negotiation would be to require them to share their private information at the start of the negotiation. However, such an approach is not feasible in the computing continuum due to the agents' self-interested nature: other agents may try to exploit an agent that reveals its private information. A better approach is to try to learn attributes of other agents (such as preferences) based on the information derived from the exchanged offers, that is, learning opponent models \cite{Baarslag2016}. Opponent modelling is an essential part of an agent's adaptation capability because it enables an agent to change its behavior based on what it believes about its opponents. Adaptation is a crucial requirement for implementing negotiation in real-life settings, as, for example, van Bragt and La Poutr\'{e} \cite{vanBragt2003} have shown that non-adaptive agents can be exploited after collecting a sufficient history of their negotiation behavior.

Baarslag et al. \cite{Baarslag2016} provide a comprehensive survey of opponent modelling in bilateral negotiation. Their main findings are that the used learning methods for opponent modelling fall generally into four categories (Bayesian learning, non-linear regression, kernel density estimation and ANN based learning), and that the models aim to learn some combination of four opponent attributes (acceptance strategy, deadline, preference profile and bidding strategy). Bayesian learning and non-linear regression are more adept for online learning as their estimates improve incrementally, whereas kernel density estimation and ANN based learning are more suitable when a training phase with sufficiently large negotiation history can be conducted.

What type of opponent modelling approaches for negotiation could be the most suitable and efficient for computing continuum is an open research question. Comparing current propositions is difficult due to lack of evaluation benchmarks for negotiation, meaning that many approaches are validated in the authors' own settings \cite{Baarslag2016}. Further, as pointed out by Baarslag et al. \cite{Baarslag2016}, many approaches make assumptions that can be very limiting. For example, non-linear regression based approaches assume that the opponent’s bidding strategy follows a known decision function with unknown parameters. In addition, opponent modelling in some settings seems to be underexplored, such as estimating an opponent’s acceptance strategy in multi-issue negotiation.

Argumentation-based negotiation enables sharing meta-information during negotiation in addition to the bids or offers \cite{rahwan2003, Rahwan2009}. The meta-information is shared in the form of arguments that try to affect the opponent's mental state. Rather than just alternating between proposals and counterproposals until a point of agreement or non-agreement is reached, argumentation tries to help the agents to understand why an offer was rejected. If an opponent rejected the offer due to not knowing some alternative way to reach its goals, an agent can try to convince the opponent to accepting the offer by providing this alternative way for reaching the goals. Argumentation can help the agents to find an agreement quicker and increase the likelihood of reaching an agreement in the first place, as well as increase the quality of the final outcome \cite{rahwan2003, Rahwan2007, Rahwan2009}.

Whether and to which extent argumentation can be utilized in the computing continuum is an open research question. Even though argumentation-based negotiation is a very compelling approach for efficient negotiation, its current research state involves many open issues that hinder its applicability in open systems such as the computing continuum. 

Argumentation aims to influence the mental state of another agent (e.g., its beliefs or goals) and requires a protocol for enabling the argumentative discussion. The knowledge state of an agent and the arguments must be expressed in a formal way in order to allow agents to reason about them. Currently, to the best of our knowledge, there are no consensus on how to represent arguments, which is crucial for enabling interoperability of agents in an open system. Further, it seems that the research on argumentation-based negotiation has strongly relied on the assumption that the agents are benevolent, without an incentive to be dishonest or withhold arguments for their own advantage. Such an assumption obviously does not hold in the computing continuum. To account for such a strategic behavior of agents in argumentation, a careful design of argumentation mechanisms is required \cite{Rahwan2009GT}.

\subsubsection{Learning to Communicate}
Communication is an essential part of a MAS. In the computing continuum, agents must exchange information with each other to ease the partial observability, make coordinated decisions and reach agreements. Enabling the communication between agents does not necessarily mean that agents must use some sort of predetermined communication protocol. Instead, a particularly interesting research question is whether the agents could learn a communication protocol while acting in the environment.

Some MADRL methods have been proposed for learning a communication protocol. Most notably, Foerster et al. \cite{Foerster2016rial} propose Reinforced Inter-Agent Learning (RIAL) and Differentiable Inter-Agent Learning (DIAL), each of which uses the CTDE paradigm and assumes a fully cooperative setting. Both methods use an ANN that outputs the agent’s Q-values for environmental actions and for communication actions. RIAL is based on Deep Recurrent Q-Networks and can also use parameter sharing, that is, a single network is learned whose parameters are used by all agents. DIAL, on the other hand, directly passes gradients via the communication channel during training, the purpose of which is to provide feedback about the communication actions and speed up the learning.

Overall, the research on learning a communication protocol in complex and open real-life settings is at scarce. A complex setting means the type of environment where the number of agents can be large, and the agents are not fully cooperative, that is, they have their own individual utilities. Such a setting would require more consideration from the agent to whom send the messages rather than just broadcasting them to everyone. Further, a mixed setting requires the inclusion of trust and reputation mechanisms, because an agent cannot blindly trust every message it gets, and agents should be incentivized to be truthful in their communication. 

An open setting means the type of environment where agents can leave and enter at any time. In the open setting, an agent needs to quickly learn the protocol the other agents have already learned. Moreover, the following questions remain open: 1) how the entry of new agents to the learning procedure will adapt the way older agents understand the messages, and 2) how older agents that learned a protocol at some point, then dropped out and joined in after the protocol had already evolved will be able to quickly adjust to the new meanings.

AI approaches for learning a protocol have also other unresolved issues that hinder their suitability for the computing continuum. They currently confine agents to using the same type of learning methods and models, which works when the MAS is implemented by the same group of developers, but not when it comes to learning protocols between agents implemented by different groups of developers. Further, they can be too resource consuming for the agents in the computing continuum, such as is the case with approaches based on using Deep Neural Networks (DNNs). Hence, alternative approaches should also be studied.

It is good to note that the traditional MAS protocols are not suitable for communication in the computing continuum, because they are specified as message flows without reference to the meaning of the messages, and they rely on a small set of primitives (i.e., message types) with unique definitions \cite[pp. 108-114]{Weiss2013}. These definitions do not usually provide enough flexibility to accommodate the communication requirements of different applications, meaning that the designers must come up with ad hoc methods to convey different types of meanings inside the messages. Consequently, agents become tightly coupled. In addition, the focus on exchanging messages in a particular order can be seen as overconstraining an agent's interactions, and thus, violating its autonomy.

Alternative solutions to agent communication in the computing continuum could come from commitment-based protocols \cite[pp. 118-121]{Fornara2004ACA, Chopra2011, Weiss2013}, where agents communicate with each other by creating and manipulating commitments. Commitment-based protocols are defined by stating the meaning of the application-specific message types in terms of the commitments that they establish between the communicating agents. These type of protocols can enforce loose coupling and autonomy of the agents, as they do not impose a specific ordering on the message flow between two parties, there is no need to hardcode into the agents how each application-specific message type should be interpreted as this interpretation can be derived from the protocol specification, and no agent is expected to provide more than it has committed to. In other words, commitments provide an underlying higher level of abstraction that can be used to provide the semantics of an application-specific message type.


\subsection{Other approaches}
\subsubsection{Generative Adversarial Networks}
Generative models can be defined as models that take a training data set drawn from some distribution, and then learn to represent an estimate of that distribution \cite{goodfellow2017nips}. Generative models can offer a lot for achieving an autonomous computing continuum. They are able to represent and manipulate high-dimensional probability distributions, which is an important ability for the computing continuum agents that are faced with high-dimensional state and action spaces. Generative models can simulate possible future states, for example, by learning a conditional distribution over future states, given the current state and possible actions. An agent can then query this model to find an action that will most likely result in a desired state (see, e.g., \cite{Finn2016}). Hence, generative models enable learning in an imaginary environment, where an agent can `imagine' the consequences of its possible actions, and mistaken actions cause no harm to the agent. Generative models can also be used for guiding exploration of an agent and for inverse reinforcement learning \cite{goodfellow2017nips, finn2016connection}.

Besides reinforcement learning related benefits, generative models are also useful in supervised learning settings. For example, they can be used to augment the training data \cite{bousmalis2017}, which is particularly beneficial in the case of ANN models that require substantial amounts of training data to generalize well. In distributed learning, they can also mitigate the non-IID problem through data augmentation \cite{jeong2018}. As another example, data synthesizing combined with self-supervised learning can help in automatically classifying service traffic data at the edge of the network without requiring any pre-knowledge of the available services or human effort for data annotation \cite{Xiao20206G}. In other words, synthesizing data can help us in reliably identifying emerging patterns, which is a crucial skill in the dynamic computing continuum.

Generative Adversarial Networks (GANs) are a type of generative models that have achieved very promising results in domains such as image generation, natural language, time-series synthesis, digital pathology and security \cite{wiatrak2020stabilizing, cai2021gansec}. The basic idea of a GAN is to set up a zero-sum game between two players \cite{goodfellow2017nips}. The first player is called a generator, which creates samples that are supposed to come from the same distribution as the training data. The second player is a discriminator that divides its input into two classes: fake or real. The ultimate goal is to reach an equilibrium, where the generator is able to create samples that are indistinguishable from the real samples.

The generator learns an implicit presentation of the density function, allowing a user to draw samples from it. GANs have several advantages over other types of generative models, such as they are not dependent on Markov chains, they can generate samples in parallel, and there is very few restrictions for the generator design. At the same time, GANs introduce a new type of challenging disadvantage: training a GAN is equal to finding an NE of a game \cite{goodfellow2017nips}.

In more detail, the generator is defined by a function that maps latent variables into observed variables. The discriminator is defined by a function that takes the observed variables as an input and outputs a single scalar value that indicates whether the input was real or fake. Both functions must be differentiable with respect to their inputs and parameters, and they are typically implemented as DNNs \cite{goodfellow2017nips}. Both the generator and the discriminator have a cost function that is dependent on their own parameters, as well as those of the other player. Because of this dependence on the parameters of both and the ability to control only one's own parameters, the cost minimization problem can be seen as a zero-sum game, the solution of which is an NE.

The main challenges plaguing the training of GANs are non-stable training, mode collapse and vanishing gradients \cite{wiatrak2020stabilizing}. Non-stable training stems from finding the equilibrium to the game between the players. While each player may reduce its loss on its individual update, that update can simultaneously cause the other player's loss to go up, resulting in a cycle where the players repeatedly undo each other's progress \cite{goodfellow2017nips}. Mode collapse refers to the phenomenon where several different latent inputs to the generator result in the same output, reducing the diversity of the generator outputs. Vanishing gradients slow down or completely stop the training of the generator. This is a major issue especially at the beginning of the training, when the discriminator is able to separate the fake inputs from the real ones with high confidence, which saturates the loss of the generator \cite{Goodfellow2014gan}.

A plethora of methods and heuristics have been proposed to mitigate at least some of the aforementioned issues in GAN training. One popular approach is to modify the loss functions of the discriminator and generator \cite{wiatrak2020stabilizing}. For example, Arjovsky et al. \cite{arjovsky2017wasserstein} proposed Wasserstein GAN (WGAN) to mitigate all three issues of non-stability, mode collapse and vanishing gradients. WGAN is based on minimizing Earth Mover's (EM) distance between the real data distribution and the generator data distribution, rather than the Jensen-Shannon divergence used in the original GAN \cite{Goodfellow2014gan}. Using EM as loss measure has shown significant improvements in the stability of the training, and it has the benefit of being a more meaningful metric, because it is a measure of distance in a space of probability distributions (i.e., it converges to zero as the distributions get close to each other) \cite{arjovsky2017wasserstein, wiatrak2020stabilizing, Hong2019gan}.

Other mitigation approaches include changing the architecture of the deep neural networks, introducing several generators or discriminators (to prevent especially the mode collapse), and modifying the optimization algorithm \cite{wiatrak2020stabilizing}. A major problem in many proposed heuristics and methods is the lack of solid theoretical foundation, which means that it depends on the context whether a particular approach is helpful \cite{goodfellow2017nips, wiatrak2020stabilizing}. For example, a study by Lucic et al. indicate that it may be more worthwhile to properly tune the hyperparameters of the models rather than turn to alternative modified methods \cite{lucic2018gan}.

Furthermore, how to objectively compare the performance of different GANs is largely an open question, as it is not undisputable nor trivial what metrics to use for the quantitative evaluation of the generated samples \cite{lucic2018gan}. It is also good to note that the majority of GAN related work has been done in image-based applications (see e.g. \cite{Hong2019gan}). Hence, applying GANs in other types of problems appearing in the computing continuum environment is an open and promising research direction.

\subsubsection{Few-shot learning}
Typically a lot of training data is required to train an ML model, particularly when it comes to training an ANN model. However, collecting such large amounts of data is burdensome, and, in complex and dynamic environments, it is practically impossible to obtain a diverse data set that would have enough samples of every possible pattern in the data. In addition, completely new patterns can emerge at any moment due to the evolving nature of the environment, such as unprecedented attack techniques in the huge volume of network traffic. Furthermore, the closer we get to the device level in the computing continuum, the more resource-constrained the agents become, meaning that they can have the capability to participate to model training only with small amounts of data. Hence, in the dynamic computing continuum, being able to quickly adapt to a new task based on a few samples and training iterations is essential.

\emph{Few-Shot Learning} (FSL) is an ML paradigm that is concerned with developing methods that are able to learn quickly from a limited number of examples. Currently, FSL has been mainly focused on supervised classification and regression problems \cite{Wang2020survey, Lu2020FSL}. There are also works that have developed FSL methods for reinforcement learning, with the aim of learning a policy based on only a few trajectories of state-action pairs \cite{Duan2017oneshot_imitation}.

In supervised ML, the core issue of FSL is the unreliable empirical risk minimizer \cite{Wang2020survey, bottou2007}. Empirical risk minimizer is the function that minimizes the expected loss over training data samples. This function can provide a good approximation of the expected risk minimizer, which is the function that minimizes the expected loss over training data distribution, but only if the number of available training samples is sufficiently large. However, as the number of available samples is small in FSL, the empirical risk minimizer is unrealiable, meaning that it may be far off from providing a good approximation. To alleviate this problem, prior knowledge must be leveraged in FSL. In their FSL survey, Wang et al. \cite{Wang2020survey} categorize FSL methods into three categories based on which aspect is improved using prior knowledge: data, model or algorithm. 

Data based methods augment the training data so that standard ML models can be used on the augmented data. Augmentation can happen by transforming data from the training data set itself, from a weakly labeled / unlabeled data set, or from similar data sets \cite{Wang2020survey}. In the scope of this article, these methods will not be elaborated further due to their ad hoc nature; they are tailored for a specific data set and cannot be easily applied to other data sets. Furthermore, existing methods are mostly concerned with data sets consisting of images \cite{Wang2020survey}.

Model based FSL methods aim to constrain the space of all model functions to a smaller function space by utilizing some prior knowledge \cite{Wang2020survey}. This prior knowledge can come from other related tasks (multitask learning) so that the some parameters are directly shared among the tasks, or parameters of different tasks are encouraged to be similar with each other through regularization. Another way is to learn so called embedding function based on prior knowledge and/or task specific knowledge. This embedding function maps each sample to a lower-dimensional space, where similar samples are closer together while dissimilar samples are more easily differentiated. A smaller function space can be constructed for this lower-dimensional space. Finally, generative modelling can be utilized in a case where the observed data is assumed to follow a certain distribution. The shape of this distribution is constrained by utilizing a prior distribution learned from other related tasks.

Algorithm based methods use prior knowledge to alter the search strategy for the parameters of the best function in the function space. This can happen in two ways: either prior knowledge (existing parameters or meta-learned parameters) offers a good initialization for the search, or prior knowledge (a set of related tasks) is used to learn a meta-learner (optimizer) that takes in the loss of the FSL learner and directly outputs the update to the task-specific parameters of the FSL learner \cite{Wang2020survey}.

In model and algorithm based FSL methods, meta-learning techniques offer an interesting \emph{learning to learn} aspect, where some generalizable information is learned across a variety of similar tasks, and then this information can be utilized in solving new tasks with only a small number of training samples. We will focus on meta-learning based FSL methods, because these methods do not require hand-crafting a learning algorithm for a specific task, and they allow a quick adaptation to new information, which is crucial for tasks in a dynamic computing continuum. Furthermore, the focus is mainly on methods that can be applied to different types of learning problems, namely to both supervised and reinforcement learning problems.

A representative example of FSL methods that utilize meta-learning techniques is the Model-Agnostic Meta-Learning (MAML) algorithm, proposed by Finn et al. \cite{Finn2017MAML}. MAML can be used with any model trained with gradient descent and can be adapted to different learning problems, such as supervised classification and regression, as well as policy-gradient reinforcement learning. MAML meta-learner does not require any additional parameters or place any constraints on the model architecture. It can also be used with a variety of loss functions. The key idea of MAML is to learn globally optimal initial model parameters across a variety of learning tasks, so that the parameters can then be quickly adapted to a new task by computing only a small number of gradient descent steps with a few training samples from that new task. 

One deficiency in MAML is that it does not consider the ambiguity in the model function, i.e., it only learns a single function. Modelling this ambiguity is particularly important for few-shot learning tasks, because the few samples in the task may not provide enough information to learn an unambiguous model even if the initial model parameters learned from the related tasks would be the best possible ones. Hence, Finn et al. extend MAML approach to probabilistic MAML \cite{Finn2018probMAML}, which learns a prior distribution over the global model parameters. As a result, multiple possible models can be sampled from the distribution before adapting to a new task with gradient descent, 
which allows to model the uncertainty in the parameters for the task.

Another deficiency in MAML is that it only learns an initialization; updating relies on SGD where the learning rate is manually set. Li et al. \cite{li2017metasgd} propose Meta-SGD algorithm that extends MAML by meta-learning the learning rate in addition to the initialization. 
Furthermore, the ANN architecture used in MAML is fixed. To this end, Elsken et al. \cite{Elsken2020NAS} propose MetaNAS that is applicable to gradient-based meta-learning FSL methods. The idea of MetaNAS is to meta-learn an ANN architecture in addition to the initialization parameters, which allows the adaptation to a task-specific ANN architecture.

Meta-learners trained over a variety of tasks may suffer from overfitting to some tasks, which results in poor generalization on new tasks that deviate significantly from the meta-training tasks. To address this issue, Jamal and Qi \cite{Jamal2019TAML} propose a Task-Agnostic Meta-Learning (TAML) algorithm, which learns unbiased initial model parameters by preventing the initial model from over-performing on any of the meta-training tasks. This prevention is done by minimizing an inequality measure over the losses of the sampled meta-training tasks. 
They build their approach on top of the MAML algorithm, but the idea of unbiased initial parameters is also applicable to many other meta-learning algorithms besides MAML \cite{Jamal2019TAML}.

Gradient-based meta-learning approaches that aim to find a shared starting point for task-specific adaptation, such as MAML, can suffer from poor generalization in high dimensional parameter spaces. This is due to the very small amount of data that is used to calculate the gradients, which often results in overfitting. Consequently, MAML is limited to using very simple, shallow ANN architectures. To mitigate this issue, Rusu et al. \cite{rusu2018leo} propose Latent Embedding Optimization (LEO) that decouples the gradient-based adaptation from the high-dimensional parameter space. This is achieved by learning a data-dependent encoder that maps input data into a low-dimensional, stochastic latent space, in which the task-specific adaptation is performed with gradient descent. A decoder, which has been learned alongside the encoder, works as a parameter generator that maps from the latent space to the high-dimensional parameter space. 


In general, FSL methods have many open issues that hinder their real-life applicability. One very fundamental issue in meta-learning or multi-task learning based FSL methods is the task-relatedness. Meta-training is conducted over thousands of batches of FSL tasks with the assumption that the meta-training tasks relate to the new tasks on a level that leads to performance improvement in the new FSL task after adaptation. Negative transfer, where the performance on a new task deteriorates after adaptation, should be avoided. With regard to reducing the number of meta-training batches, it may be worthwhile to study how to combine transfer learning from pre-trained DNN models with meta-learning \cite{sun2018metatransfer}.

Majority of FSL related work has been done in the field of image classification \cite{Wang2020survey, Lu2020FSL, Hospedales2021metasurvey}, and the performance of these methods is far behind the traditional supervised methods \cite{sun2018metatransfer, Hospedales2021metasurvey}. FSL methods usually also rely on unrealistic assumptions, such as that the samples and their labels in the support set for the FSL training are accurate and reliable, and that there exist a huge, properly labeled data set from which to extract prior knowledge for the FSL task. Studying FSL more in the context of unsupervised and semi-supervised learning would be beneficial for real-life applications (see e.g. \cite{xu2021unsupFSL, Lu2020FSL}). 

FSL methods for supervised classification often assume that the model for the FSL task needs to only reason about the novel classes in the support set and not about any previously seen classes. This phenomenon is referred to as catastrophic forgetting \cite{Lu2020FSL}. Especially in computing continuum, where new class concepts can appear in a dynamic manner, forgetting the old classes should not happen. For example, a security model trained to classify network traffic should not forget about previous attack classes when it is adapted to recognize new ones.

Many of the existing FSL methods also assume that the input-output data distribution in the FSL tasks is the same as in the previously seen tasks and data sets. This is an unrealistic assumption in dynamic environments where the conditions under which data is produced can change, resulting in shifts in the input-output distribution. These kind of situations require methods that can account for cross-domain adaptation \cite{Adler2020CrossDomainFL, Lu2020FSL}.

Finally, how to combine FSL with distributed learning in the computing continuum environment is a significant and interesting research direction with regard to our vision, because assuming that each agent in a computing continuum system would have thousands of labeled samples for collaborative model training is unrealistic. Some preliminary work that combines FSL with FL has been done. For example, 
Fan and Huang \cite{Fan2021FFSL} propose an adversarial learning procedure for federated FSL in supervised classification settings. The method aims to find an initialization for a global model consisting of a feature encoder and a classifier so that it quickly adapts to a new FSL task with unseen classes. However, their adversarial learning method can be computationally too heavy for resource-constrained participants, and their experiments only consider up to 30 participants, while not accounting for, e.g., stragglers or domain shift (i.e., the data for training and testing comes from the same data set in their experiments).

The combination of FSL and distributed learning is very uncharted. Preliminary work considers only supervised classification in a centralized FL setting with a small number of participants. A lot more research is required to assess the performance, scalability and efficiency of FSL in a variety of distributed learning settings, as well as the effects of a large number of heterogeneous participants, non-IID data, domain shifts, and possible decentralization of the learning. 

\subsubsection{Self-supervised learning}
Supervised learning requires that every sample has a corresponding label. Obtaining a labeled data set for ANN training can require an expensive and time-consuming manual data labeling process, which usually also limits the size of the data set. In computing continuum, where the training data is distributed among several nodes, it is unrealistic to assume that each node has a properly labeled, high-quality data set with substantial number of samples. Furthermore, the question of who is annotating the data in a geographically distributed computing continuum is not a trivial question. Any extensive attempts to manually label data in such an environment are already made infeasible by the enormous amount of data that a distributed, dynamic system can generate every day at a fast pace. Hence, an essential question arises: how can the large amounts of unlabeled data be utilized in model training?

\emph{Self-supervised learning} (SSL) is a learning paradigm that tries to tackle the data annotation issue by utilizing some type of automatic supervision signal extracted from the unlabeled data. This usually means learning a representation from the unlabeled data (often called a pretext task) that can be subsequently used in a model trained with a smaller amount of labeled data (often called a downstream task). SSL has been most commonly studied in the context of representation learning for image based and natural language processing applications, and more recently in an increasing amount for applications using graph data \cite{Liu2021self, liu2022GSSL, Wu2021GSSL}.

Liu et al. \cite{Liu2021self} categorize self-supervised representational learning into three categories: generative, contrastive and adversarial. The core idea behind each class of methods is to obtain an excellent pre-trained feature extractor for the downstream task, which requires designing a proper, downstream task related objective for the pretext task. Generative SSL methods use reconstruction loss to train a encoder-decoder architecture, contrastive methods use a a similarity metric based, contrastive loss in the latent space to train an encoder, and adversarial methods use a loss based on distributional divergence to train a generator for creating fake samples and a discriminator for distinguishing them from the real ones.

Generative methods have been popular in classification and generation because they are able to recover the original data distribution without assumptions for downstream tasks \cite{Liu2021self}. However, these methods have been inferior to contrastive methods in image classification. In addition, the reconstruction loss is often based on maximum likelihood, which means that the data distribution is modelled on a point-wise level (e.g., pixels in images), which is a low-level abstraction and makes the model very sensitive to rare samples.

Contrastive methods assume that the downstream applications are classification tasks. They have been very successful in image classification tasks, but their suitability for classification tasks in other domains such as natural language processing has not been extensively studied \cite{Liu2021self}. Furthermore, many constrastive methods require so called negative sampling, because they are often based on distinguishing between representations coming from augmentations of the same data point (positive samples) and those of other data points (negative samples). What type of negative sampling scheme to use is an essential question, and the role and necessity of negative sampling in contrastive methods is an open issue \cite{Liu2021self, Tian2021SSL}.

Adversarial methods are mainly based on GANs. However, GANs only learn an implicit latent representation and thus, cannot be directly used in self-supervised representation learning. Some methods that aim to utilize GANs in representational learning try to extract the latent representation by replacing the generator with an adversarial autoencoder \cite{Makhzani2016AAE} or by treating the generator as a decoder and adding an encoder to learn a mapping from real samples to the latent space (i.e., a converted generator) \cite{Dumoulin2017AdversariallyLI}. Another way to use GANs for representational learning is to train them to recover the whole output based on a partial input (e.g., image completion \cite{Iizuka2017inpainting}) rather than training them to recover the whole output based on a latent input. GANs have been mainly utilized in vision based applications \cite{Liu2021self}. Extending them to a wider variety of self-supervised applications is an interesting research direction due to their ability to model high-level abstractions.

It is good to note that self-supervision does not remove the need for labeled data, but it can help to reach better performance compared to supervised learning when a small amount of labeled data is available. Newell et al. have shown in vision based tasks that SSL is not able to reach higher accuracy than supervised learning when there is enough labeled data available, but it can help to reach a better accuracy in low data regimes \cite{Newell2020self}. Even though SSL cannot help in improving performance beyond supervised methods with large amounts of labeled data, it still can help in model robustness and uncertainty estimation, as shown by Hendrycks et al. \cite{Hendrycks2019}. They find that SSL can make models more robust to adversarial examples, label corruption, and common input image corruptions, as well as better at detecting out-of-distribution samples.

As noted, SSL has been mainly studied in representational learning for image classification in centralized training settings. Applying SSL in a wider variety of problems appearing in the computing continuum environment is an important research direction, because acquiring large amounts of properly labeled data in a large distributed system is practically impossible. Applying SSL in computing continuum requires research on how to effectively enable distributed SSL. Some preliminary work that combine SSL with FL exist (see, e.g., \cite{vanBerlo2020FSSL, Saeed2021FSSL, zhuang2022divergenceaware}). Most notably, Saeed et al. \cite{Saeed2021FSSL} propose an approach that uses wavelet transform to learn representations from unlabeled sensor inputs. Their pretext task is a contrastive method where a deep temporal neural network is trained to determine if a given pair of a signal and its complementary view 
align with each other. 
For training this network in a distributed manner, they adopt a vanilla FL setting, assuming a small number of participants at each training round and dividing the training data across participants randomly. Subsequently, more research on distributed SSL is required to assess, among other things, the effects of a large number of heterogeneous participants with limited resources, non-IID data, and complete decentralization of the training.

There are many other challenges besides the distribution of self-supervised training that hinder the use of SSL methods in autonomous computing continuum, some of the major ones being the following: establishing theory that could help in avoiding empirical misconceptions and adapting solutions from vision based domains to other domains; automating the design of pretext tasks, which in its current form seems to be done in a very ad hoc manner and relies heavily on human effort; and model adaptation when data distribution shifts.

\section{Edge AI for Continuum Orchestration}\label{sec:Discussion}
\begin{table}[]
    \centering
    \caption{Potential edge AI uses in continuum orchestration.}
    \begin{tabular}{l p{0.28\textwidth}}
    \toprule
         \textbf{Orchestration function} & \textbf{Edge AI usage}  \\
    \midrule
        Monitoring & Distributed learning (\cref{sec:distlearning}) for nowcasting, forecasting and simulating, e.g., node workloads or user mobility. \\
         \midrule
        Dataflow management & Decision making (\cref{sec:decmaking}) on aggregation, sharing, offloading, caching (e.g., optimal timing). Distributed learning (\cref{sec:distlearning}) techniques to train ML models to support decision making. \\ 
         \midrule
        Discovery & Negotiation (\cref{sec:negotiation}) for resources across and within administrative boundaries.\\
         \midrule
        Allocation & Decision making (\cref{sec:decmaking}) on placement, scheduling, migration, scaling, replication (e.g., optimal timing). Distributed learning techniques (\cref{sec:distlearning}) to train ML models to support decision making. \\ 
         \midrule
        Lifecycle management & Decision making (\cref{sec:decmaking}) on starting, stopping, updating (e.g., optimal timing).\\
    \bottomrule
    \end{tabular}
    \label{tab:edgeai_orchestration}
\end{table}

Our vision models the computing continuum as a MAS consisting of autonomous, intelligent and self-interested agents. The focal point of the vision is that the computing continuum agents must make intelligent decisions regarding the orchestration functions (see  \cref{fig:orchestration-taxonomy}). In this section, we briefly illustrate how edge AI can be useful in different orchestration functions. \cref{tab:edgeai_orchestration} summarizes the potential edge AI usage for each of the functions.

\textit{Monitoring.} Monitoring, performance estimation and benchmarking processes could especially benefit from ML models that can give accurate short-term and long-term predictions about the changing conditions in the environment. This capability would support pre-emptive adaptivity in the system. Further, ML models can help the monitoring function to estimate the current status of a node/resource of which there is no first-hand information available. Moreover, inside one administrative domain, the agents responsible for monitoring the system could deploy a collaborative peer-to-peer learning approach to train the models (\cref{sec:distlearning}).

\textit{Dataflow management.} Distributed learning (\cref{sec:distlearning}) and decision making (\cref{sec:decmaking}) techniques are very useful for dataflow management. For example, an agent can learn an intelligent decision making strategy about whether the workload should be executed on the node or sent to another node. This decision can be based on, e.g., the current state of the node, network conditions, and predictions about changes in the workload. ML models can support the decision making by providing the required predictions. 

\textit{Discovery.} Discovery could mainly benefit from negotiation methods (\cref{sec:negotiation}). For example, agents in charge of the discovery function could negotiate in a competitive setting across administrative boundaries for the prices of additional resources. These resource could then be made available across the domains. Furthermore, within administrative boundaries, agents in different clusters, with differing service level sub-objectives (see \cref{sec:synthesis} and \cref{fig:vision-objective}), may negotiate in a co-operative setting for resource usage. 

\textit{Allocation.} Similarly to dataflow management, distributed learning (\cref{sec:distlearning}) and decision making (\cref{sec:decmaking}) techniques are very useful for allocation. Intelligent decision making strategies can make optimized decisions about, e.g., what is the optimal time to migrate a container or replicate a service. ML models can support the decision making through nowcasting or forecasting.

\textit{Lifecycle management.} Decision making techniques (\cref{sec:decmaking}) can help in making timely decisions about starting, stopping or updating services. 

\section{Conclusion}\label{sec:Conclusion}
In this article, we studied orchestration in the device-edge-cloud continuum, and focused on \textit{AI for edge}, that is, AI methods used in the orchestration of the resources in the device-edge-cloud computing continuum. We claimed that to support the constantly growing requirements of intelligent applications in the computing continuum, resource orchestration needs to embrace edge AI and emphasize local autonomy and intelligence. 

To justify the claim, we provided a general definition for continuum orchestration, and looked at how current and emerging orchestration paradigms are suitable for the computing continuum. We described certain major emerging research themes that may affect future orchestration, and provided an early vision of an orchestration paradigm that embraces those research themes. 

Finally, we surveyed current key edge AI methods and looked at how they may contribute into fulfilling the vision of future continuum orchestration. We identified methods for distributed machine learning, decision making and negotiation, described their fundamentals, and discussed their strengths and weaknesses in relation to the challenges in the computing continuum. Finally, we reflected upon edge AI methods in relation to the functionality required of continuum orchestration, presenting potential uses for AI capabilities in implementing the orchestration functions.

\section*{Acknowledgment}

The authors would like to thank the researchers and alumni at the Center for Ubiquitous Computing (UBICOMP) at the University of Oulu, and Distributed Systems Group (DSG) at TU Wien, for the discussions and insights during the writing of this manuscript.
\printbibliography

\end{document}